\documentclass[preprint,aps,prd,superscriptaddress]{revtex4-1}
\usepackage{graphicx}
\usepackage{slashed}
\usepackage{verbatim} 
\newcommand{\be}{\begin{eqnarray}}
\newcommand{\ee}{\end{eqnarray}}
\usepackage[colorlinks]{hyperref}
\usepackage{cancel,bm} 
\usepackage{amsmath,amssymb}
\usepackage{mathtools}
\usepackage{float}
\usepackage{color}
\usepackage{leftidx}
\usepackage{booktabs}
\usepackage{latexsym}
\usepackage{revsymb}
\usepackage{multirow}
\usepackage{hypcap}
\usepackage{array}
\usepackage[english]{babel}
\usepackage{subfigure}
\newcommand{\xB }{x_{\scriptscriptstyle B}}
\newcommand{\sT}{{\scriptscriptstyle T}}

\begin{document}

\title{\texorpdfstring{$\cos2\phi_t$} ~~azimuthal asymmetry in  back-to-back \texorpdfstring{$J/\psi$}~~-jet production in   \texorpdfstring{$e ~p\rightarrow e ~J/\psi ~Jet~ X$}~ ~at the EIC}

\author{Raj Kishore}
\email{raj.theps@gmail.com}
\affiliation{Department of Physics, Indian Institute of Technology Kanpur, Kanpur-208016, India}
\author{Asmita Mukherjee}
\email{asmita@phy.iitb.ac.in}
\affiliation{Department of Physics, Indian Institute of Technology Bombay, Mumbai-400076, India}
\author{Amol Pawar}
\email{194120018@iitb.ac.in}
\affiliation{Department of Physics, Indian Institute of Technology Bombay, Mumbai-400076, India}
\author{Mariyah Siddiqah}
\email{shah.siddiqah@gmail.com}
\affiliation{Department of Physics, Indian Institute of Technology Bombay, Mumbai-400076, India}
\date{\today}

\begin{abstract}
In this article, we investigate the $\cos2\phi_t$ azimuthal asymmetry in   $e ~p\rightarrow e ~J/\psi ~Jet~ X$, where the $J/\psi$-jet pair is almost back-to-back in the transverse plane, within the framework of the generalized parton model(GPM). We use non-relativistic QCD(NRQCD) to calculate the $J/\psi$ production amplitude and incorporate both color singlet(CS)  and color octet(CO) contributions to the asymmetry. We estimate the asymmetry using different parameterizations of the gluon TMDs in the kinematics that can be accessed at the future electron-ion collider (EIC) and also investigate the impact of transverse momentum dependent (TMD) evolution on the asymmetry. We present the contributions coming from different states to the asymmetry in NRQCD.
\end{abstract}

\maketitle
\raggedbottom 

\section{Introduction}
Transverse momentum dependent parton distributions (TMDs)\cite{Mulders:1995dh,boer1998time,boer2000angular,anselmino1999phenomenology,anselmino1995single,barone2002transverse} give a tomographic picture of the nucleon in terms of quarks and gluons in momentum space. TMDs play an important role in processes where two scales are involved; for example, in semi-inclusive deep inelastic scattering (SIDIS) where apart from the 
photon virtuality, one measures the transverse momentum of the outgoing particle, or Drell-Yan process where the transverse momentum of the outgoing lepton-pair provide the second scale. For such processes, one can apply the generalized factorization involving TMDs. However, the TMD factorization has not been proven for all processes. In the kinematical limit when the collinear factorization becomes valid, the result based on TMD factorization should be matched with that obtained using  collinear factorization in the same process, with the inclusion of soft factors in the TMD framework \cite{collins2011foundations,Echevarria:2011epo,echevarria2013soft,Echevarria:2019ynx,Bacchetta:2008xw,Fleming:2019pzj}. To ensure gauge invariance, the TMDs include gauge links or Wilson lines, which comes due to initial or final state
interactions \cite{collins2002leading,ji2002parton,belitsky2003final,
boer2003universality}. These  introduce process dependence in them. Recent results from RHIC are in favour of the theoretical prediction of a sign change of the Sivers function observed in SIDIS and DY processes, respectively\cite{Anselmino:2016uie}.
More data is needed to have a firm understanding of the process dependence of the TMDs.  Gluon TMDs \cite{mulders2001transverse} till now are far less investigated than the quark TMDs. The positivity bound gives a constraint on them \cite{mulders2001transverse}.  Recently, there is an extraction of unpolarized gluon TMDs from LHCb data \cite{lansberg2018pinning}. Gauge invariance of the gluon TMDs requires the inclusion of two gauge links in the definition, which makes the process dependence more involved than the quark TMDs \cite{buffing2013generalized}. The most common are one past and one future pointing gauge links, $[+-]$ or $[-+]$ (f-type) and both past or both future pointing, $[--]$ or $[++]$ (d-type). The operator structures of these two types of gluon TMDs are different \cite{buffing2013generalized}. In the literature on small-$x$ physics, these two gluon TMDs are called Weiszacker-Williams (WW) \cite{kovchegov1998gluon,mclerran1999fock} type and dipole \cite{dominguez2012linearly} type, respectively. These contribute in different processes. 

In an unpolarized proton, there is a nonzero probability of finding linearly polarized gluons. The linearly polarized gluon distributions were introduced in \cite{mulders2001transverse}, and first investigated in a model in \cite{meissner2007relations}. Recently these have attracted quite a lot of attention, although till now they have not been extracted using data. Linearly polarized gluon distributions can be probed in $ep$ 
and $pp$ collisions \cite{Marquet:2017xwy,Pisano:2013cya,boer2009dijet,efremov2018measure,efremov2018ratio,lansberg2017associated,dumitru2019measuring,sun2011gluon,boer2013determining,boer2012linearly,Echevarria:2015uaa,boer2012polarized,mukherjee2017linearly,mukherjee2016probing}.
These give an azimuthal asymmetry of the form $\cos~2 \phi_t$\cite{Pisano:2013cya}; also, they affect the transverse momentum distribution of the outgoing particle. Depending on the gauge links, the linearly polarized gluon distribution can be WW or dipole type. These are time reversal even (T-even) objects. In $pp$ scattering processes, the initial and final state interactions  often affect the TMD factorization; the asymmetries and cross sections also involve both f-type and d-type gluon TMDs, and disentangling the two is difficult from the observables. The gauge link structure is simpler in $ep$ scattering processes \cite{Rajesh:2021zvd}, and the upcoming electron-ion collider (EIC) at Brookhaven National Lab will play an important role in probing the gluon TMDs, including the linearly polarized gluon TMD over a wide kinematical region. 

$\cos~2 \phi_t$ asymmetry in $J/\psi$ production in unpolarized $ep$ collision has been  shown to be a useful observable to probe the linearly polarized gluon TMD. Contribution to the asymmetry comes already at the leading order (LO) through the virtual photon-gluon fusion process \cite{mukherjee2017j}; this contributes at $z=1$, where $z$ is the fraction of the energy of the photon carried by the $J/\psi$ in the rest frame of the proton. In the kinematical region $z<1$ \cite{kishore2019accessing} one has to incorporate higher order Feynman diagrams. In this process, the $J/\psi$ produced needs to be detected in the forward region, or with its transverse momentum $\mathrm{p}_T$ not so large; otherwise TMD factorization is not expected to hold. In this work, we investigate the $\cos2 \phi_t$ asymmetry in a slightly different process, namely when a $J/\psi$ and a jet are observed almost back-to-back in $ep$ collision. Only the WW type gluon TMDs contribute in this process \cite{DAlesio:2019qpk}. Here the $J/\psi$ produced can have large transverse momentum, as the soft scale required for the TMD factorization is provided by  the total transverse momentum of the $J/\psi$-jet pair, which is smaller than their invariant mass as they are almost back-to-back \cite{Kishore:2019fzb}. In fact, by varying the invariant mass of the pair, one can also probe the TMDs over a wide range of scales and investigate  the effect of TMD evolution on the asymmetry. In \cite{DAlesio:2019qpk}, the upper bound of the $\cos~2 \phi_t$ asymmetry was investigated in this process, as well as the asymmetry in the small-$x$ region. Here we present a calculation of the asymmetry using some recent parameterization of the gluon TMDs and also investigate the effect of TMD evolution.  

A widely used approach to calculate the amplitude of $J/\psi$ production is based on an effective field theory called non-relativistic QCD (NRQCD) \cite{hagler2001towards,yuan2001polarizations,yuan2008heavy}.
Here, one assumes that the amplitude for $J/\psi$ production process can be factorized into a hard part where the $c\bar{c}$ pair is produced perturbatively, and a soft part where the heavy quark pair hadronizes to form a $J/\psi$. The hadronization process is encoded in the long distance matrix elements (LDMEs) \cite{Bodwin:1994jh}; which are usually extracted using the data. The cross-section is expressed as a double expansion in terms of the strong coupling $\alpha_s$ as well as the velocity parameter associated with the heavy quark $v$\cite{Boer:2021ehu,lepage1992improved};
in the limit $v\ll1$. For charmonium $v \approx 0.3$. The heavy quark pair  in the hard process is produced in different states denoted by $^{2s+1} L_{J}^{(c)}$ where $s$ denotes the spin of the pair (singlet or triplet), $L$ is the orbital angular momentum, $J$ is the total angular momentum and $(c)$ denotes the color configuration, which can be singlet or octet. The heavy quark pair produced in the hard process emits soft gluons to evolve into $J/\psi$. For the $S$-wave contribution, the dominant term in the limit $v \approx 0$ gives the result of the color singlet model (CSM)\cite{berger1981inelastic,baier1983hadronic}, where the heavy quark pair in the hard process is assumed to be produced with the same quantum numbers as the $J/\psi$, and in the color singlet state. In our work, we include both color singlet (CS) and color octet (CO) contributions.  

 The paper is arranged as follows: In section II, we present the TMD formalism  adopted.  In Section  III, we investigate the effect of the TMD evolution on the asymmetries. In sections IV and V, we present two recent parameterizations of the gluon TMDs, based on the spectator model and a Gaussian parameterization, respectively. Numerical results are  presented in section VI. The conclusion is discussed  in section VII.
 
\section{Formalism}
We consider a semi-inclusive electroproduction of a $J/\psi$ and a jet, 
\begin{equation}
e^-(l) + p(\mathrm{P}) \to e^- (l') + J/\psi (\mathrm{P}_{\psi}) + Jet (\mathrm{P}_{j}) + X, \nonumber
\end{equation} 
 where the four-momentum of the particles are given in their corresponding round brackets.  Here, both the incoming electron beam and the target proton are unpolarized with their respective momenta $l$ and $P$. The kinematics of the process can be described in the following variables
\begin{eqnarray}
Q^2=-q^2, ~~~~~~ s=(P+l)^2, ~~~~~~~~~ W^2=(P+q)^2,\\
x_B=\frac{Q^2}{2P\cdot q}, ~~~~~~~~~~ y=\frac{P\cdot q}{P\cdot l},~~~~~~~~~~~~~z=\frac{P\cdot \mathrm{P}_{\psi}}{P\cdot q}.
\end{eqnarray}
The virtuality of the scattering photon is given by $Q^2$, $s$ is the square of the electron-proton center of mass energy whereas, $W$ is the invariant mass of photon-proton system. $x_B$ is the Bjorken-$x$ variable, $y$ is the inelasticity variable that gives the fraction of the energy of the electron taken by the scattering virtual photon and the variable $z$ defines the fraction of energy of the photon carried by the outgoing $J/\psi$ particle in proton rest frame. We consider virtual photon-proton center of mass frame where they move along $+z$ and $-z$ directions, respectively. To define the kinematics, we use the light-cone coordinate system. We use two light-like vectors, one of which is chosen to be in the direction of the proton momentum, $P=n_{-}$ and the other $n=n_{+}$, such that $P\cdot n=1$ and $n_{-}^2=n_{+}^2=0$. In terms of these vectors, the momenta of the particles involved in the process can be written as follows;\\
Momenta of the initial proton and virtual photon  can be expressed as  
\begin{eqnarray}
P^{\mu}&&=n_{-}^{\mu}+\frac{M_p^2}{2}n_{+}^{\mu}\approx n_{-}^{\mu},\\
q^{\mu}&&=-x_B n_{-}^{\mu}+\frac{Q^2}{2x_B}n_{+}^{\mu}\approx -x_B P^{\mu}+(P\cdot q)n_{+}^{\mu},
\end{eqnarray}
where $M_p$ is proton mass and $Q^2=x_Bys$. The expressions for the incoming and outgoing lepton momenta can be written in terms of light cone coordinates using the inelasticity variable $y$ as,
\begin{eqnarray}
l^\mu&& = \frac{(1-y)x_B}{y} P^\mu + \frac{(P\cdot q)}{y} n^\mu + \frac{\sqrt{1-y}}{y}Q\hat{l}_{\perp}^{\mu},\\
l'^\mu&&=l^\mu-q^\mu.
\end{eqnarray}
At the partonic level process, $J/\psi$ can either be produced through the gluonic channel: $g+\gamma^\star\to J/\psi +g$ or through the quark (anti-quark) channel: $q(\bar{q})+\gamma^\star\to J/\psi+q(\bar{q})$. However, in the small-$x$ region, the gluonic channel dominates over the quark (anti-quark) channel \cite{DAlesio:2019qpk}. Hence, we have considered the contribution from the gluonic channel only in the following estimate of the asymmetry. The above  processes contribute at the next-to-leading (NLO) order in $\alpha_s$ and in the kinematic region  $z<1$. The outgoing energetic gluons produced in this process gives the jet. Now, we can express the momentum of the initial gluon, $p_g$, the final $J/\psi$ and jet with momentum $\mathrm{P}_\psi$ and $\mathrm{P}_j$, respectively, in terms of the light-like vectors as
%
\begin{eqnarray}
&&p_g^\mu = xP^\mu +(p_g\cdot P - M_p^2 x)n^\mu + \textbf{p}_T^\mu \approx xP^\mu + \textbf{p}_T^\mu,\\ 
&&\mathrm{P}_{\psi}^\mu =\frac{\textbf{P}_{\psi\perp}^2 +M_\psi^2}{2zP\cdot q}P^\mu + z(P\cdot q )n^\mu + \textbf{P}_{\psi\perp}^\mu,\\
&&\mathrm{P}_{j}^\mu = \frac{\textbf{P}_{j\perp}^2}{2(1-z)P\cdot q}P^\mu + (1-z)(P\cdot q )n^\mu +\textbf{P}_{j\perp}^\mu, \label{4p}
\end{eqnarray}
where, $x$ is the collinear momentum fraction of the initial gluon, 
$\textbf{P}_{\psi\perp}$ and $\textbf{P}_{j\perp} $ are the transverse momenta of $J/\psi$ and jet respectively.
The incoming and the outgoing scattered lepton form the  leptonic plane. All the azimuthal angles of the final state particles are defined with respect to the leptonic plane with $\phi_l=\phi_{l'}=0$. 

For the process under consideration, TMD factorization has not formally been proven yet, although it is expected to be valid. In our study we have assumed TMD factorization. The total differential scattering cross-section for the  unpolarized process: $ep\to J/\psi\ Jet\ X$ can be written as \cite{Pisano:2013cya}
\begin{eqnarray}
 d\sigma &=& \frac{1}{2s}\frac{d^3l'}{(2\pi)^32E_{l'}}\frac{d^3\mathrm{P}_{\psi}}{2E_{\psi}(2\pi)^3}\frac{d^3\mathrm{P}_{j}}{2E_{j}(2\pi)^3} \int dx~d^2\textbf{p}_{T}~ (2\pi)^4 \delta^4(q+p_g-\mathrm{P}_{j}-\mathrm{P}_{\psi})\nonumber\\
 && \frac{1}{Q^4}~ L^{\mu\mu'}(l,q)~~\Phi^{\nu\nu'}_{g}(x,\textbf{p}_T^2)~ ~\mathcal{M}_{\mu\nu}^{g\gamma^*\to J/\psi\ g} ~~\mathcal{M}_{\mu'\nu'}^{*g\gamma^*\to J/\psi\ g}. \label{totsig}
 \end{eqnarray} 
 The function $\mathcal{M}_{\mu\nu}$ represents  the scattering amplitude of $J/\psi$ production in  the photon-gluon fusion process: $\gamma^*(q) + g (p_g) \rightarrow Q\bar{Q}(\mathrm{P}_{\psi}) + g(\mathrm{P}_{j})$
partonic subprocess. The leptonic tensor $L^{\mu\mu'}$  describes the electron-photon scattering and can be written as,
\begin{eqnarray}
 L^{\mu\mu'}&=e^2(-g^{\mu\mu'}Q^2+2(l^{\mu}l^{'\mu'}+l^{\mu'}l^{'\mu})),
\end{eqnarray}
where $e$ is represents the electronic charge. The gluon correlator, $\Phi^{\nu\nu'}_{g}(x,\textbf{p}_T^2)$ describes the gluon content of the proton. At the leading twist, for the case of an unpolarized proton, it can be parameterized in terms of two TMD gluon distribution functions as \cite{mulders2001transverse}
\begin{eqnarray}
 \Phi_g^{\nu\nu'}(x,\mathbf{p}_{T}^2)=-\frac{1}{2x}\bigg\{g_{\perp}^{\nu\nu'}f_1^g(x,\mathbf{p}_{T}^2)-\left(\frac{p_{T}^{\nu}p_{T}^{\nu'}}{M_p^2}+g_{\perp}^{\nu\nu'}\frac{\mathbf{p}_{T}^2}{2M_p^2}\right)h_1^{\perp g}(x,\mathbf{p}_{T}^2)\bigg\}.
\end{eqnarray}
Here, $g_{\perp}^{\nu\nu'}=g^{\nu\nu'}-P^{\nu}n^{\nu'}/P\cdot n-P^{\nu'}n^{\nu}/P\cdot n$. The quantities $f_1^g(x,\mathbf{p}_{T}^2) $ and $h_1^{\perp g}(x,\mathbf{p}_{T}^2)$ represent the unpolarized and  linearly polarized gluon TMD, respectively.
\subsection{\texorpdfstring{$J/\psi$} ~~production in NRQCD framework}
\begin{figure}[H]
    \includegraphics[width=16.5cm,height=8.4cm]{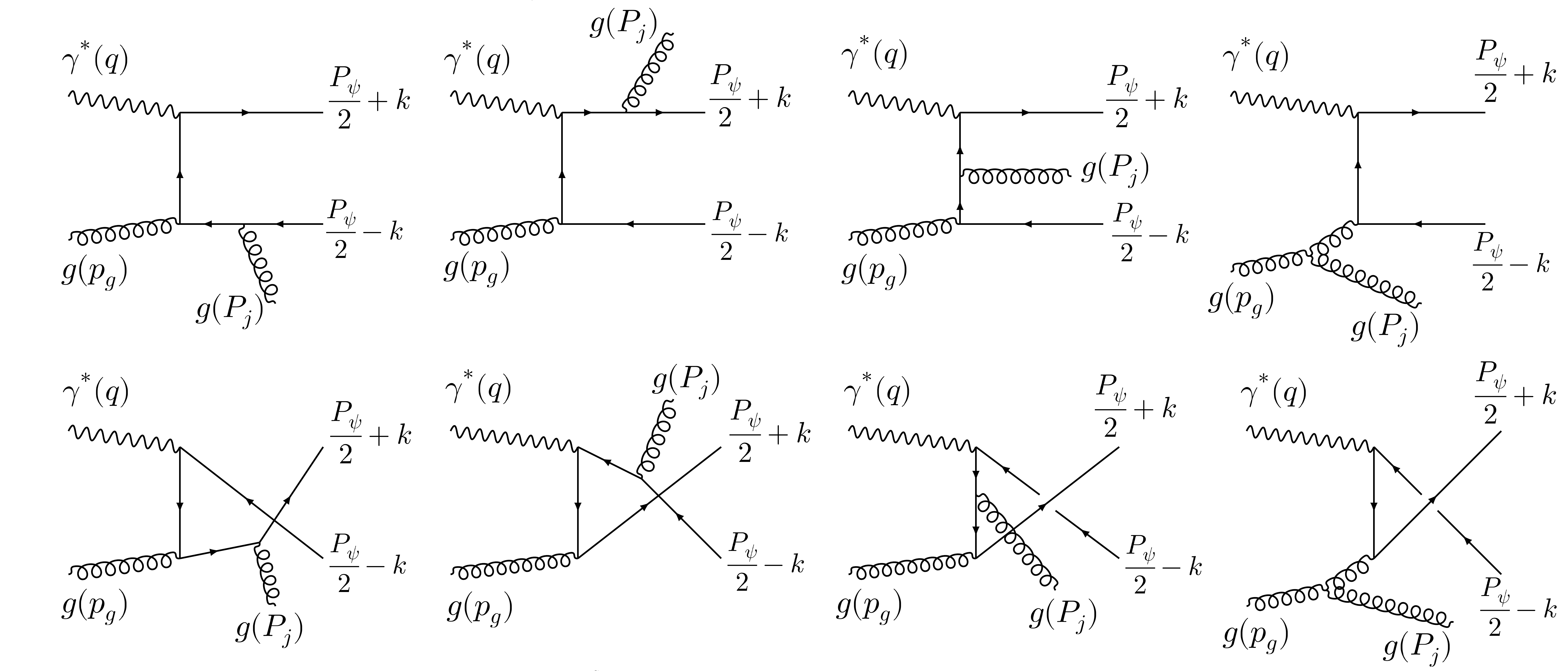}
   \caption{  Feynman diagrams for the partonic process $\gamma^*(q) + g (p_g) \rightarrow J/\psi(\mathrm{P}_{\psi}) + g(\mathrm{P}_{j})$.}\label{figFeyndiag}
 \end{figure}

The Feynman diagrams for the dominant subprocess of photon-gluon fusion, which results in the production of a $J/\psi$ and a jet, are shown in Fig.~\ref{figFeyndiag}.
The amplitude for the production of $J/\psi$ within the NRQCD framework can be written as follows \cite{boer2012polarized,baier1983hadronic}
\begin{eqnarray}
\mathcal{M}\Big(\gamma^*~g \to Q\bar{Q}[^{2S+1}L_{J}^{(1,8)}]~g \Big)=\sum_{L_zS_z}&& \int \frac{d^3\textbf{k}}{(2\pi)^3} \Psi_{LL_z}(\textbf{k}) \langle LL_z;SS_z|JJ_z \rangle \nonumber\\
&&\times Tr[O(q,p_g,\mathrm{P}_{\psi},k)\bm{\mathcal{P}}_{SS_z}(\mathrm{P}_{\psi},k)],\label{eqnmatrix_ele} 
\end{eqnarray}
where $k$ is the relative momentum of the heavy quark or the anti-quark in the rest frame of the non-relativistic quarkonium bound state, which is assumed to be very small as compared with the rest mass of the quarkonium. Here $\Psi_{LL_z}(\textbf{k})$ is the nonrelativistic
bound-state wave function with orbital angular momentum  $L,L_z$. The Clebsch-Gordan coefficients $\langle LL_z;SS_z|JJ_z \rangle$ projects out their angular momentum. The mass of the quarkonium, $M_\psi$, is taken to be twice the heavy quark mass.     
The $O(q,p_g,\mathrm{P}_{\psi},k)$ represents the amplitude for the
production of the heavy quark anti-quark pair, $Q\bar{Q}$, without the inclusion of the polarization of the quark and anti-quark. This can be calculated by considering the contributions from all the above Feynman diagrams and can be written as
\begin{equation}
O(q,p_g,\mathrm{P}_{\psi},k)=\sum_{i=1}^8 C_i~O_i(q,p_g,\mathrm{P}_{\psi},k), \label{eqnO_matrixele}
\end{equation}   
where, $i$ denotes the contribution from the individual Feynman diagrams given in Fig. \ref{figFeyndiag} and $C_i$ corresponds to the color factor for each diagram. The $O_i(q,p_g,\mathrm{P}_{\psi},k)$ for the above Feynman diagrams are written as
\begin{eqnarray}
O_1&=&4g_s^2(ee_c)\varepsilon_g^{*\lambda}(\mathrm{P}_j)\gamma_\nu \frac{\slashed{\mathrm{P}}_{\psi}+2\slashed{k}-2\slashed{q}+M_\psi}{(\mathrm{P}_{\psi}+2k-2q)^2-M_\psi^2}\gamma_\mu \frac{-\slashed{\mathrm{P}}_{\psi}+2\slashed{k}-2\slashed{\mathrm{P}}_{j}+M_\psi}{(\mathrm{P}_{\psi}-2k+2\mathrm{P}_{j})^2-M_\psi^2}\gamma_\lambda,\nonumber\\
O_2&=&4g_s^2(ee_c)\varepsilon_g^{*\lambda}(\mathrm{P}_j)\gamma_\lambda \frac{\slashed{\mathrm{P}}_{\psi}+2\slashed{k}+2\slashed{\mathrm{P}}_{j}+M_\psi}{(\mathrm{P}_{\psi}+2k+2\mathrm{P}_{j})^2-M_\psi^2}\gamma_\nu \frac{-\slashed{\mathrm{P}}_{\psi}+2\slashed{k}+2\slashed{p}_g+M_\psi}{(\mathrm{P}_{\psi}-2k-2p_g)^2-M_\psi^2}\gamma_\mu ,\nonumber\\
O_3&=&4g_s^2(ee_c)\varepsilon_g^{*\lambda}(\mathrm{P}_j)\gamma_\nu \frac{\slashed{\mathrm{P}}_{\psi}+2\slashed{k}-2\slashed{q}+M_\psi}{(\mathrm{P}_{\psi}+2k-2q)^2-M_\psi^2}\gamma_\lambda \frac{-\slashed{\mathrm{P}}_{\psi}+2\slashed{k}+2\slashed{p}_g+M_\psi}{(\mathrm{P}_{\psi}-2k-2p_g)^2-M_\psi^2}\gamma_\mu,\nonumber\\
O_4&=&4g_s^2(ee_c)\varepsilon_g^{*\lambda}(\mathrm{P}_j)\gamma_\nu \frac{\slashed{\mathrm{P}}_{\psi}+2\slashed{k}-2\slashed{q}+M_\psi}{(\mathrm{P}_{\psi}+2k-2q)^2-M_\psi^2}\gamma^{\chi}\nonumber\\
&&~~~~~~~~~~~~~~~~~~~~~~~~~~~\frac{[g_{\mu\lambda}(p_g+\mathrm{P}_j)_{\chi}+g_{\lambda\chi}(p_g-2\mathrm{P}_j)_{\mu}+g_{\chi\mu}(\mathrm{P}_j-2p_g)_{\lambda}]}{(p_g-\mathrm{P}_j)^2}.\label{eqnmatele}
\end{eqnarray}
The expressions for the remaining, $O_5,~O_6,~O_7$ and $O_8$ 
can be obtained by reversing the fermionic current and replacing $k$ to $-k$.

In the NRQCD framework, the outgoing $Q\bar{Q}$ pair can be formed in the color singlet (CS) state or in the color octet (CO) states. The color factors, $C_i$, corresponding to the CO case are given as \cite{rajeshCO},
\begin{eqnarray}
&&C_1=C_6=C_7=\sum_{jk} \langle 3j;\bar{3}k|8c \rangle (t_at_b)_{jk}, \nonumber\\
&&C_2=C_3=C_5=\sum_{jk} \langle 3j;\bar{3}k|8c \rangle (t_bt_a)_{jk}, \nonumber\\ 
&&C_4=C_8=\sum_{jk} \langle 3j;\bar{3}k|8c \rangle if_{abd} (t_d)_{jk}. \label{eqcolorfac1}
\end{eqnarray}
The SU(3) Clebsch-Gordon coefficients for CS and CO states are given by, respectively,
\begin{equation}
\langle 3j;\bar{3}k|1 \rangle = \frac{\delta_{jk}}{\sqrt{N_c}},~~\langle 3j;\bar{3}k|8c \rangle = \sqrt{2}(t_c)_{jk},\label{sU3clebsch}
\end{equation}
where $N_c$ is number of colors, $t_c$ is the generators of SU(3) in the fundamental representation. Their properties are given by $Tr[t_at_b]=\delta_{ab}/2$ and $Tr[t_at_bt_c] =\frac{1}{4}(d_{abc}+if_{abc})$. By using these relations along with the relation in Eq. (\ref{sU3clebsch}), one obtains the color factors for individual Feynman diagrams for the $Q\bar{Q}$ formed in color octet states as
\begin{eqnarray}
&&C_1=C_6=C_7=\frac{\sqrt{2}}{4}(d_{abc}+if_{abc}),\nonumber\\
&&C_2=C_3=C_5=\frac{\sqrt{2}}{4}(d_{abc}-if_{abc}),\nonumber\\
&&C_4=C_8=\frac{\sqrt{2}}{2}if_{abc}.\label{colorfactorco}
\end{eqnarray}
For the case of formation of $Q\bar{Q}$ pair in CS state, the color factors are given by
\begin{equation}\label{colorfactorcs}
    C_1=C_2=C_3=C_5=C_6=C_7=\frac{\delta_{ab}}{2\sqrt{N_c}}.
\end{equation}
The spin projection operator for the bound state of the $J/\psi$ includes the spinors of the heavy quark and anti-quark, $c\bar{c}$ and is given as
\begin{eqnarray}
\bm{\mathcal{P}}_{SS_z}(\mathrm{P}_{\psi},k)&&=\sum_{s_1s_2} \Big{\langle}\frac{1}{2}s_1;\frac{1}{2}s_2 \Big{|}SS_z\Big{\rangle} v\Big(\frac{\mathrm{P}_{\psi}}{2} - k,s_1\Big)\bar{u}\Big(\frac{\mathrm{P}_{\psi}}{2} + k,s_2\Big)\nonumber\\
&&=\frac{1}{4M_\psi^{3/2}} \big(-\slashed{\mathrm{P}}_{\psi}+2\slashed{k}+M_\psi\big)\Pi_{SS_z}\big(\slashed{\mathrm{P}}_{\psi}+2\slashed{k}+M_\psi\big)+\mathcal{O}(k^2),\label{eqnSpinpolari}
\end{eqnarray} 
where, $\Pi_{SS_z} = \gamma^5$ for spin singlet ($S=0$) and $\Pi_{SS_z} =\slashed{\varepsilon}_{S_z}(\mathrm{P}_{\psi}) $ for spin triplet ($S=1$). The $\varepsilon_{S_z}$ is the spin polarization vector of the outgoing $c\bar{c}$ pair.

Now, since, in the rest frame of the bound state,  $k \ll \mathrm{P}_{\psi}$, thus one can Taylor expand the amplitude given in Eq. (\ref{eqnmatrix_ele}) around $k=0$ limit. In that expansion, the terms corresponding to $k^0$ give S-wave scattering amplitude $(L=0,~J=0,1)$ and terms linear in $k$ correspond to P-wave scattering $(L=1,~J=0,1,2)$ and the corresponding amplitudes are given as
\begin{eqnarray}
\mathcal{M}[^{2S+1}S_{J}^{(1,8)}](\mathrm{P}_{\psi},k)=&& \frac{1}{\sqrt{4\pi}}R_0(0)Tr[O(q,p_g,\mathrm{P}_{\psi},k)\bm{\mathcal{P}}_{SS_z}(\mathrm{P}_{\psi},k)]\Big|_{k=0}\nonumber\\
=&&\frac{1}{\sqrt{4\pi}}R_0(0)Tr[O(0)\bm{\mathcal{P}}_{SS_z}(0)],\label{sstate}\\
\mathcal{M}[^{2S+1}P_{J}^{(8)}](\mathrm{P}_{\psi},k)=&&-i\sqrt{\frac{3}{4\pi}}R_1'(0)\sum_{L_zS_z}\varepsilon^{\alpha}_{L_z}(\mathrm{P}_\psi)\langle LL_z;SS_z|JJ_z\rangle\nonumber\\
&&\times\frac{\partial}{\partial k^\alpha}Tr[O(q,p_g,\mathrm{P}_{\psi},k)\bm{\mathcal{P}}_{SS_z}(\mathrm{P}_{\psi},k)]\Big|_{k=0}\nonumber\\
=&&-i\sqrt{\frac{3}{4\pi}}R_1'(0)\sum_{L_zS_z}\varepsilon^{\alpha}_{L_z}(\mathrm{P}_\psi)\langle LL_z;SS_z|JJ_z\rangle\nonumber\\ &&\times Tr[O_\alpha(0)\bm{\mathcal{P}}_{SS_z}(0)+O(0)\bm{\mathcal{P}}_{SS_z\alpha}(0)]. \label{Pstate}
\end{eqnarray}
Where $R_L$ represents the radial wave function and the shorthand notations used in the above expressions are  
\begin{eqnarray}
&&O(0)=O(q,p_g,\mathrm{P}_{\psi},k)\Big|_{k=0},~~~~~~~~~~~~~\bm{\mathcal{P}}_{SS_z}(0)=\bm{\mathcal{P}}_{SS_z}(\mathrm{P}_{\psi},k)\Big|_{k=0},\\
&&O_\alpha(0)=\frac{\partial}{\partial k^\alpha}O(q,p_g,\mathrm{P}_{\psi},k)\Big|_{k=0}, ~~~~~~\bm{\mathcal{P}}_{SS_z\alpha}(0)=\frac{\partial}{\partial k^\alpha}\bm{\mathcal{P}}_{SS_z}(\mathrm{P}_{\psi},k)\Big|_{k=0}.
\end{eqnarray}\\
Following are the relations for the Clebsch Gordon coefficient and the polarization vector of $J/\psi$ which we can use to calculate $P$ wave amplitudes \cite{Pisano2012}
\begin{eqnarray}
&&\sum_{L_zS_z}\langle LL_z;SS_z|JJ_z\rangle \varepsilon^\alpha_{S_z}(\mathrm{P}_\psi)\varepsilon^\beta_{L_z}(\mathrm{P}_\psi)=\sqrt{\frac{1}{3}}\Bigg(g^{\alpha\beta}-\frac{\mathrm{P}_\psi^\alpha\mathrm{P}_\psi^\beta}{M_\psi^2} \Bigg),\\
&&\sum_{L_zS_z}\langle LL_z;SS_z|JJ_z\rangle \varepsilon^\alpha_{S_z}(\mathrm{P}_\psi)\varepsilon^\beta_{L_z}(\mathrm{P}_\psi)=-\frac{i}{M_\psi}\sqrt{\frac{1}{2}}\epsilon_{\delta\zeta\xi\varrho}g^{\xi\alpha}g^{\varrho\beta}\mathrm{P}_\psi^\delta\varepsilon^\zeta_{J_z}(\mathrm{P}_\psi),\\
&&\sum_{L_zS_z}\langle LL_z;SS_z|JJ_z\rangle \varepsilon^\alpha_{S_z}(\mathrm{P}_\psi)\varepsilon^\beta_{L_z}(\mathrm{P}_\psi)=\varepsilon^{\alpha\beta}_{J_z}(\mathrm{P}_\psi).
\end{eqnarray}
The $\varepsilon^\alpha_{J_z}(\mathrm{P}_\psi) $ is the polarization vector corresponding to $J=1$ angular momentum state, that follows the current conservation and obeys the following relations \cite{Pisano2012},
\begin{eqnarray}
\varepsilon^\alpha_{J_z}(\mathrm{P}_\psi)\mathrm{P}_{\psi\alpha}&=&0,\\
\sum_{J_z}\varepsilon^\alpha_{J_z}(\mathrm{P}_\psi)\varepsilon^{*\beta}_{J_z}(\mathrm{P}_\psi)&=&\Bigg(-g^{\alpha\beta}+\frac{\mathrm{P}_\psi^\alpha\mathrm{P}_\psi^\beta}{M_\psi^2}\Bigg)=\mathcal{Q}^{\alpha\beta}.
\end{eqnarray}
The $\varepsilon^{\alpha\beta}_{J_z}(\mathrm{P}_\psi)$ is the polarization tensor corresponding to $J=2$ which is symmetric in the Lorentz indices and follows the relations \cite{Pisano2012},
\begin{eqnarray}
\varepsilon^{\alpha\beta}_{J_z}(\mathrm{P}_\psi)=\varepsilon^{\beta\alpha}_{J_z}(\mathrm{P}_\psi)~\varepsilon^\alpha_{J_z\alpha}(\mathrm{P}_\psi)=0~\varepsilon^\alpha_{J_z}(\mathrm{P}_\psi)\mathrm{P}_{\psi\alpha}=0,\nonumber\\
\varepsilon^{\alpha\beta}_{J_z}(\mathrm{P}_\psi)\varepsilon^{*\mu\nu}_{J_z}(\mathrm{P}_\psi)=\frac{1}{2}[\mathcal{Q}^{\alpha\mu}\mathcal{Q}^{\beta\nu}+\mathcal{Q}^{\alpha\nu}\mathcal{Q}^{\beta\mu}]-\frac{1}{3}[\mathcal{Q}^{\alpha\beta}\mathcal{Q}^{\mu\nu}].
\end{eqnarray}
The radial wave function and its derivative evaluated at origin $R_0(0)$, $R_1'(0)$ given in  Eq. (\ref{sstate}) and Eq. (\ref{Pstate})  are related to LDMEs by the following equations \cite{DAlesio:2019qpk},
\begin{eqnarray}
&&\langle0|\mathcal{O}_1^{J/\psi}(^{2S+1}S_J)|0\rangle=\frac{N_c}{2\pi}(2J+1)|R_0(0)|^2,\\
&&\langle0|\mathcal{O}_8^{J/\psi}(^{2S+1}S_J)|0\rangle=\frac{2}{\pi}(2J+1)|R_0(0)|^2,\\
&&\langle0|\mathcal{O}_8^{J/\psi}(^{3}P_J)|0\rangle=\frac{2N_c}{\pi}(2J+1)|R'_1(0)|^2.
\end{eqnarray}
The two different sets of LDMEs that we have used in our numerical calculation are given in Table. \ref{tabLDME}
 \begin{table}[H]
 \centering
 \begin{tabular}[c]{c|c|c|c|c|c}
 
 \hline \hline
                   &    $\langle0|\mathcal{O}_8^{J/\psi}(^{1}S_0)|0\rangle$             & $\langle0|\mathcal{O}_8^{J/\psi}(^{3}S_1)|0\rangle$              &$\langle0|\mathcal{O}_1^{J/\psi}(^{3}S_1)|0\rangle$  &$\langle0|\mathcal{O}_8^{J/\psi}(^{3}P_0)|0\rangle/m_c^2$&\\
 \hline			
Ref. \cite{SharmaLDME} & 1.8 $\pm$ 0.87 &0.13 $\pm $0.13 & 1.2 $\times~10^2$ & 1.8 $\pm$ 0.87 & $\times10^{-2}$GeV$^3$                 \\
Ref. \cite{ChaoLDME} & 8.9 $\pm$ 0.98 &0.30 $\pm $0.12 & 1.2 $\times~10^2$ & 0.56 $\pm$ 0.21 & $\times10^{-2}$GeV$^3$     \\
 \hline \hline
  \end{tabular}
   \caption{Numerical values for two different sets of LDMEs.}
 \label{tabLDME}
 \end{table}
 Now, using the aforementioned formalism, together with the symmetry relations among the amplitudes corresponding to different Feynman diagrams for each state, which have been calculated in Ref.~\cite{rajesh2018sivers},  we could write the amplitude for the   CS state $(^3S_1^{(1)})$ and CO states $(^3S_1^{(8)},\ ^1S_0^{(8)},\ ^3P_{J(0,1,2)}^{(8)} )$ as
 \subsection{\texorpdfstring{$^3S_1^{(1,8)}$} ~~Amplitude}
 The final expression for  $^3S_1^{(1)} $state and $^3S_1^{(8)}$ state can be written as,
 \begin{eqnarray}
  \mathcal{M}[^{3}S_{1}^{(1)}](\mathrm{P}_{\psi},p_g)=\frac{1}{4\sqrt{\pi M_\psi}}R_0(0)\frac{\delta_{ab}}{2\sqrt{N_c}}Tr\Bigg[\sum_{i=1}^3 O_i(0)(\slashed{\mathrm{P}}_{\psi} +M_\psi)\slashed{\varepsilon}_{S_z}\Bigg] \label{CSMA},\\
   \mathcal{M}[^{3}S_{1}^{(8)}](\mathrm{P}_{\psi},p_g)=\frac{1}{4\sqrt{\pi M_\psi}}R_0(0)\frac{\sqrt{2}}{2}d_{abc}Tr\Bigg[\sum_{i=1}^3 O_i(0)(\slashed{\mathrm{P}}_{\psi} +M_\psi)\slashed{\varepsilon}_{S_z}\Bigg] \label{COMA}.
 \end{eqnarray}
 Where, $\sum_{i=1}^3 O_i(0) $ is given as,
 \begin{eqnarray}
 \sum_{i=1}^3O_i(0)&&=4g_s^2(ee_c)\varepsilon_g^{*\lambda}\Bigg[ \frac{\gamma_\nu(\slashed{\mathrm{P}}_{\psi}-2\slashed{q}+M_\psi)\gamma_\mu(-\slashed{\mathrm{P}}_{\psi}-2\slashed{\mathrm{P}}_{j}+M_\psi)\gamma_\lambda}{(s-M_\psi^2)(u-M_\psi^2+q^2)}\nonumber\\
&&~~~~~~~~~~~~~~~~~~~~+\frac{\gamma_\lambda(\slashed{\mathrm{P}}_{\psi}+2\slashed{\mathrm{P}}_{j}+M_\psi)\gamma_\nu(-\slashed{\mathrm{P}}_{\psi}+2\slashed{p}_g+M_\psi)\gamma_\mu}{(s-M_\psi^2)(t-M_\psi^2)}\nonumber\\ &&~~~~~~~~~~~~~~~~~~~~+\frac{\gamma_\nu(\slashed{\mathrm{P}}_{\psi}-2\slashed{q}+M_\psi)\gamma_\lambda (-\slashed{\mathrm{P}}_{\psi}+2\slashed{p}_g+M_\psi)\gamma_\mu }{(t-M_\psi^2)(u-M_\psi^2+q^2)} \Bigg].\nonumber\\ \label{total3S1}
 \end{eqnarray}
The symmetry relations, given in Ref.~\cite{rajesh2018sivers}, lead to the cancellation of contributions from Feynman diagrams 4 and 8 for  $^3S_1^{(1,8)}$ amplitude.
 
  \subsection{\texorpdfstring{$^1S_0^{(8)}$} ~~Amplitude}
  The total amplitude for $^1S_0^{(8)}$ state can be written as,
  \begin{equation}
  \mathcal{M}[^{1}S_{0}^{(8)}](\mathrm{P}_{\psi},p_g)=\frac{1}{4\sqrt{\pi M_\psi}}R_0(0)i\frac{\sqrt{2}}{2}f_{abc}Tr\Big[(O_1(0)-O_2(0)-O_3(0)+2O_4(0))(\slashed{\mathrm{P}}_{\psi} +M_\psi)\gamma_5\Big].       \label{amp1S0}
  \end{equation}
  $O_1(0),~O_2(0),~O_3(0)$ are given in the Eq. (\ref{total3S1}) and  
  \begin{equation}
 O_4(0)=    g_s^2(ee_c)\varepsilon_g^{*\lambda} \frac{\gamma_\nu(\slashed{\mathrm{P}}_{\psi}-2\slashed{q}+M_\psi)\gamma^{\chi}}{u(u-M_\psi^2)}[g_{\mu\lambda}(p_g+\mathrm{P}_j)_{\chi}+g_{\lambda\chi}(p_g-2\mathrm{P}_j)_{\mu}+g_{\chi\mu}(\mathrm{P}_j-2p_g)_{\lambda}]. 
  \end{equation}
    \subsection{\texorpdfstring{$^3P_J^{(8)}$ }~~Amplitude}
 The $^3P_J^{(8)}$ amplitude can be written as  \cite{rajesh2018sivers},
 \begin{eqnarray}
 \mathcal{M}[^{3}\mathcal{P}_{J}^{(8)}](\mathrm{P}_{\psi},p_g)=&&\frac{\sqrt{2}}{2}f_{abc}\sqrt{\frac{3}{4\pi}}R_1'(0)\sum_{L_zS_z}\varepsilon^{\alpha}_{L_z}(\mathrm{P}_\psi)\langle LL_z;SS_z|JJ_z\rangle\nonumber\\ &&Tr[(O_{1\alpha}(0)-O_{2\alpha}(0)-O_{3\alpha}(0)+2O_{4\alpha}(0))\bm{\mathcal{P}}_{SS_z}(0)\nonumber\\
 &&+(O_1(0)-O_2(0)-O_3(0)+2O_4(0))\bm{\mathcal{P}}_{SS_z\alpha}(0)].
 \end{eqnarray}

\subsection{Total differential cross-section and the asymmetry }
\begin{figure}[H]
  \begin{center}
  \includegraphics[width=14cm,height=14cm,keepaspectratio]{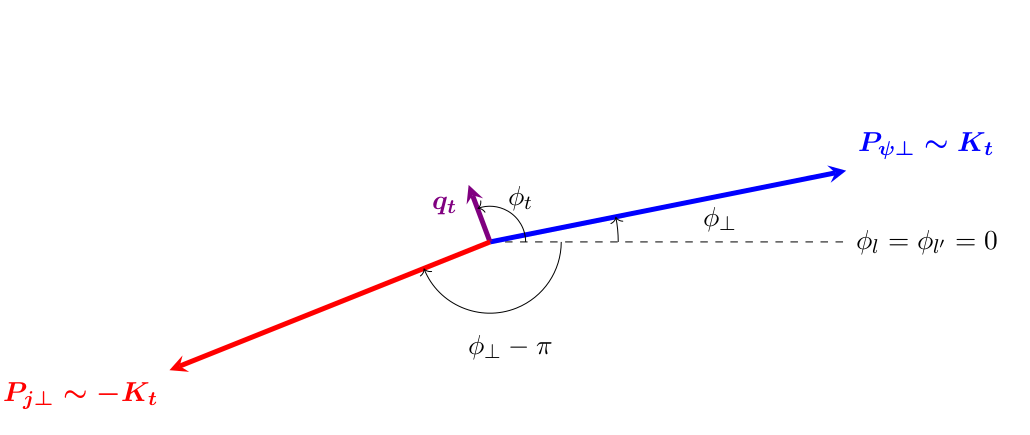}
  \caption{The schematic representation of back to back $J/\psi$ and jet production in transverse plane.  }
  \label{figbtb}
  \end{center}
\end{figure}

The structure of differential cross-section as defined in Eq. (\ref{totsig}) has a contraction of tensors which can be schematically written as
\begin{equation}
\mathfrak{M}_i\mathfrak{M}_j = L^{\mu\mu'}(l,q) \Phi_g^{\nu\nu'}(x,\mathbf{p}_{T})\mathcal{M}_{i\mu\nu}\mathcal{M}_{j\mu'\nu'},
\end{equation}
where $i,j=1,2,3,4$ (corresponds to the Feynman diagrams given in Fig.~\ref{figFeyndiag} and we have $\mathfrak{M}_i\mathfrak{M}_j = \mathfrak{M}_j\mathfrak{M}_i$ for $ i\neq j$. We have already defined all tensors in the above convolution and contributions to the amplitudes come from all the CS and CO states $(^3S_1^{(1,8)},\ ^1S_0^{(8)},\ ^3P_{J(0,1,2)}^{(8)} )$. We have summed over all the polarization states of the outgoing gluon using the relation given as
\begin{eqnarray}
\sum_{\lambda_a = 1}^{2} \varepsilon^{\lambda_a}_{\mu}\varepsilon^{\lambda_a}_{\mu'}= -g_{\mu\mu'}+\frac{\mathrm{P}_{j\mu}n_{\psi\mu'}+\mathrm{P}_{j\mu'}n_{\psi\mu}}{\mathrm{P}_{j}\cdot n_{\psi}}-\frac{\mathrm{P}_{j\mu}\mathrm{P}_{j\mu'}}{(\mathrm{P}_{j}\cdot n_\psi)^2},\label{eqnpol_final_gluon}
\end{eqnarray}
where, $n_{\psi\mu}=\mathrm{P}_{\psi\mu}/M_\psi$.
We use  the frame where the incoming virtual photon and proton moves along $z$-axis. The azimuthal angles of the lepton scattering plane is defined as $\phi_l=\phi_{l'}=0$. We integrate out the azimuthal angle of the final lepton $\ell'$ ~\cite{graudenz1994next}, we can write
 \begin{equation}
 \frac{\mathrm{d}^{3}\ell'}{(2\pi)^{3}2E_{\ell'}}  =\frac{\mathrm{d}Q^{2}\mathrm{d}y}{16\pi^{2}}.
 \end{equation}
 Moreover, for the other phase factors, one can write
 \begin{equation}
\frac{\mathrm{d}^{3}\mathrm{P}_{\psi}}{(2\pi)^{3}2E_{\psi}}  =\frac{\mathrm{d}z\mathrm{d}^{2}\bm{\mathrm{P}}_{\psi\perp}}{(2\pi)^{3}2z}\;,\quad\frac{\mathrm{d}^{3}\mathrm{P}_{j}}{(2\pi)^{3}2E_{j}}=\frac{\mathrm{d}\bar{z}\mathrm{d}^{2}\bm{\mathrm{P}}_{j\perp}}{(2\pi)^{3}2\bar{z}}\;,
 \end{equation}
and conservation of the four-momenta can be written as
 \begin{equation}
 \begin{aligned} & \delta^{4}\big(q+p_g-\mathrm{P}_{\psi}-\mathrm{P}_{j}\big)\\
  & =\frac{2}{ys}\delta\bigl(1-z-\bar{z}\bigr)\delta\left(x-\frac{\bar{z}(M^2+\bm{\mathrm{P}}_{\psi\perp}^2)+z \bm{\mathrm{P}}_{j\perp}^2 + z \bar{z}Q^2}{z(1-z)ys}\right )\delta^{2}\bigl(\bm{\mathrm{p}}_{\sT}-\bm{\mathrm{P}}_{j\perp}-\bm{\mathrm{P}}_{\psi\perp}\bigr)\;,
 \label{eqdelta}
 \end{aligned}
 \end{equation}
 where we have used the relation $Q^{2}=x_Bys$. Now, we define the sum and difference of the transverse momentum of $J/\psi$ and jet as 
 \begin{equation}
 \begin{aligned}\bm{\mathrm{q}}_{t} & \equiv \bm{\mathrm{P}}_{\psi\perp}+\bm{\mathrm{P}}_{j\perp}\,,\quad \bm{\mathrm{K}}_{t}\equiv\frac{\bm{\mathrm{P}}_{\psi\perp}-\bm{\mathrm{P}}_{j\perp}}{2}\,.\label{eqtransverse_kine}\end{aligned}
 \end{equation}
 From Eq. (\ref{eqdelta}), we have $\bar{z}=(1-z)$ and $\textbf{q}_t=\textbf{p}_T$. We use TMD factorization  in the kinematic region  $|\textbf{q}_t| \ll |\textbf{K}_t|$. This leads to the situation where the outgoing $J/\psi$ and jet are  almost back to back in the transverse plane, $i.e.$ the $xy$ plane,  w.r.t the virtual photon-proton colliding axis, $i.e.$ the $z$-axis, thus allowing us to set $\textbf{K}_t\simeq \textbf{P}_{\psi\perp} \simeq -\textbf{P}_{j\perp}$. $\phi_t$ and $\phi_{\perp}$ are the azimuthal angles, respectively, for $\textbf{q}_t$ and $\textbf{K}_t$ and are defined with respect to the leptonic plane as illustrated in Fig.~\ref{figbtb}. 
Finally, integrating over $\bar{z}$, $\bm{\mathrm{p}}_{_T}$ and $x$, the differential cross-section, in Eq. (\ref{totsig}), as a function of $z,\ y,\ x_B,\ \bm{\mathrm{q}}_t,\ \bm{\mathrm{K}}_t$ can be rewritten as
 \begin{equation}
 \begin{aligned}\frac{\mathrm{d}\sigma}{\mathrm{d}z\,\mathrm{d}y\,\mathrm{d}\xB \,\mathrm{d}^{2}\bm{\mathrm{q}}_{t}\mathrm{d}^{2}\bm{\mathrm{K}}_{t}} & =\frac{1}{(2\pi)^{4}}\frac{1}{16sz(1-z)Q^{4}}\sum_{i,j}\mathfrak{M}_i\mathfrak{M}_j.\end{aligned}
 \label{eq:csdetail}
 \end{equation}
%
In the following calculation, we have only kept the zeroth and first order terms  in $\Big(\frac{\bm{\mathrm{q}}_t^2}{M_P^2}\Big)$. Thus, we obtain the total differential cross-section \cite{DAlesio:2019qpk},
\begin{eqnarray}\label{eqfinaldsigma}
\frac{\mathrm{d}\sigma}{\mathrm{d}z\,\mathrm{d}y\,\mathrm{d}\xB\,\mathrm{d}^{2}\bm{\mathrm{q}}_{t} \,\mathrm{d}^{2}\bm{\mathrm{K}}_{t}} &&=\frac{1}{(2\pi)^{4}}\frac{1}{16sz(1-z)Q^{4}} \Big\{ \big(\mathbb{A}_0+\mathbb{A}_1\cos\phi_\perp + \mathbb{A}_2\cos2\phi_\perp\big)f_1^g(x,\bm{\mathrm{q}}_t^2)+
\nonumber\\
&&\frac{\bm{\mathrm{q}}_t^2}{M_P^2}h_1^{\perp g}(x,\bm{\mathrm{q}}_t^2)\big(\mathbb{B}_0\cos2\phi_t+\mathbb{B}_1\cos(2\phi_t-\phi_\perp)+\mathbb{B}_2\cos2(\phi_t-\phi_\perp)+\nonumber\\ 
&&\mathbb{B}_3\cos(2\phi_t-3\phi_\perp)+\mathbb{B}_4\cos(2\phi_t-4\phi_\perp)\big) \Big\}.
  \end{eqnarray}
The analytic expressions of the coefficients $\mathbb{A}_i$'s and $\mathbb{B}_i$'s are very lengthy, which
we have not included in this article. They are available upon request.

The TMDs come with various azimuthal modulations. These modulations can be used to extract information about the ratio of TMDs. This can be done by defining the asymmetries  as \cite{DAlesio:2019qpk} 
\begin{equation}
A^{\mathcal{W}(\phi_S,\phi_t)}\equiv 2\frac{\int\mathrm{d}\phi_S ~\mathrm{d}\phi_t \mathrm{d}\phi_\perp \mathcal{W}(\phi_S,\phi_t)~\mathrm{d}\sigma(\phi_S,\phi_t,\phi_\perp)}{\int\mathrm{d}\phi_S~ \mathrm{d}\phi_t~ \mathrm{d}\phi_\perp~ \mathrm{d}\sigma(\phi_S,\phi_t,\phi_\perp)}.
\end{equation}
Here, we are interested in one particular asymmetry, i.e.,  $\cos2\phi_t$ asymmetry to extract the linearly polarized gluon TMD. This can be written as,
\begin{equation}
\langle\cos2\phi_t\rangle\equiv A^{\cos2\phi_t} = 2\frac{\int \mathrm{d}\phi_t\mathrm{d}\phi_\perp \cos2\phi_t \mathrm{d}\sigma(\phi_t,\phi_\perp)}{\int \mathrm{d}\phi_t\mathrm{d}\phi_\perp\mathrm{d}\sigma(\phi_t,\phi_\perp)}.
\end{equation}
Now, by plugging the differential scattering cross-section from  Eq. (\ref{eqfinaldsigma}) in the above equation, we get the $\cos2\phi_t$ asymmetry as function of $z,\ y,\ x_B, \mathrm{K}_t$ 
\begin{eqnarray}
\langle\cos2\phi_t\rangle\equiv A^{\cos2\phi_t}=\frac{\int \mathrm{q}_t~ \mathrm{d}\mathrm{q}_t~\frac{\textbf{q}_t^2}{M_p^2}~\mathbb{B}_0~h_1^{\perp g}(x,\textbf{q}_t^2)}{\int \mathrm{q}_t~\mathrm{d}\mathrm{q}_t~\mathbb{A}_0~f_1^g(x,\textbf{q}_t^2)}.\label{cos2phi}
\end{eqnarray}
For estimating the $\cos2\phi_t$ asymmetry numerically, we have used two recent parameterizations of the gluon TMDs, and also estimated the effect of TMD evolution. The following section gives the details of the TMD evolution formalism used.

\section{ TMD evolution}
The evolution of the TMDs \cite{Aybat:2011zv,Aybat:2011ta} with the scale affects the asymmetries measured in the energies of different experiments, and it is important to estimate the effect of this evolution 
to obtain the angular asymmetries of produced
hadrons measured by the HERMES, COMPASS, JLab as well as the future EIC  experiments at different energies. The TMD evolution is usually studied in the impact parameter space \cite{Aybat:2011zv}.  The impact parameter dependent TMDs can be written as  Fourier transforms of the TMDs,
\begin{eqnarray}
\hat{f}(x,\bm{\mathrm{b}}_t^2;Q_f^2)&=& \frac{1}{2\pi} \int d^2\bm{\mathrm{q}}_t~ e^{i\bm{\mathrm{q}}_t\cdot\bm{\mathrm{b}}_t}~f(x,\bm{\mathrm{q}}_t^2,Q_f^2).\label{TMDBT}
\end{eqnarray}
In the TMD evolution approach, TMDs not only evolve with the intrinsic transverse momentum of the parton but also evolve with the probing scale. The expression for TMD evolution at a given final scale $Q_f$ can be obtained by solving Collins-Soper evolution equation and renormalization group equation. Using this approach, the expression for the gluon TMD in the impact parameter space can be written as \cite{echevarria2014qcd,aybat2012qcd,collins2011foundations}
\begin{equation}
\hat{f}(x,\bm{\mathrm{b}}_t^2,Q_f^2) =  \frac{1}{2\pi} \sum_{p=q,\bar{q},g} (C_{g/p}\otimes f^p_1)(x,Q_i^2)e^{-\frac{1}{2}S_A(\bm{\mathrm{b}}_t^2,Q_f^2,Q_i^2)} e^{-S_{np}(\bm{\mathrm{b}}_t^2,Q_f^2)},\label{TMDevol1}
\end{equation} 
where  $Q_i$ is the initial scale of TMD, defined in terms of $\mathrm{b}_t$ as, $Q_i=2e^{-\gamma_{E}}/\mathrm{b}_t$ and $\gamma_E\approx 0.577$. 
$C_{g/p}$ are coefficient functions and $f^p_1(x,Q^2)$ are collinear parton distributions for a species of partons like quark/anti-quark or gluon. The exponents $S_A$ and $S_{np}$ are the  perturbative and non-perturbative Sudakov factors, respectively. We note that
the Sudakov factor $S_A$ is spin independent, and thus the same
for all (un)polarized TMDs \cite{Echevarria:2015uaa,echevarria2014unified}.  As stated before, a formal proof of TMD factorization  for this process remains to be done, and here we study the effect of TMD evolution on the asymmetry assuming such factorization.
The subsections below contain a description of the TMD evolution formalism used and the relevant formulas.
\subsection{Coefficient functions and perturbative Sudakov factor}
The coefficient function  can be written as a series of strong coupling constant $\alpha_s$\cite{boer2020j},
\begin{equation}
C_{g/p}(x,Q_i)=\delta_{gp}\delta(1-x) + \sum_{k=1}^{\infty}\sum_{p=g,q,\bar{q}}C^{k}_{g/p}(x)\Bigg(\frac{\alpha_s(Q_i)}{\pi}\Bigg)^{k}.\label{Coeff}
\end{equation} 

 The Sudakov factor as the leading order of $\alpha_s$ can be written as\cite{boer2020j},
\begin{eqnarray}
S_A(\bm{\mathrm{b}}_t^2,Q_f^2,Q_i^2) &=& \frac{C_A}{\pi} \int^{Q_f^2}_{Q_i^2} \frac{d\eta^2}{\eta^2} \alpha_s(\eta) \Bigg(\log\frac{Q_f^2}{\eta^2} - \frac{11-2n_f/C_A}{6}\Bigg)\nonumber\\
&=& \frac{C_A}{\pi} \alpha_s \Bigg(\frac{1}{2}~\log^2\frac{Q_f^2}{Q_i^2}-\frac{11-2n_f/C_A}{6}~\log\frac{Q_f^2}{Q_i^2}\Bigg).\label{Suda}
\end{eqnarray}
The running of the coupling  $\alpha_s$ is ignored in Eq. (\ref{Suda}) because it starts at $\alpha_s^2$. The $C_A=N_c$, $n_f$ denotes the number of active flavours. 
In $\mathrm{b}_t \ll 1/\Lambda_{QCD}$ the Sudakov factor can be Taylor expanded. Further, by substituting the expressions from Eq. \eqref{Coeff} and Eq. \eqref{Suda} the unpolarized gluon TMD at LO without non-perturbative Sudakov factor is given as \cite{boer2020j},
\begin{eqnarray}
\hat{f}_1^g(x,\bm{\mathrm{b}}_t^2;Q_f^2) =&& \frac{1}{2\pi} \Bigg\{f_1^g(x,Q_i^2)-\frac{\alpha_s}{2\pi}\Bigg(\Bigg[\frac{C_A}{2}~\log^2\frac{Q_f^2}{Q_i^2}-\frac{11-2n_f/C_A}{6}~\log\frac{Q_f^2}{Q_i^2}\Bigg]f^g_1(x,Q_i^2)\nonumber\\
&&~~~~~~~~~~~~~~~~~~~~~~~~~~~~~~~~~~~~~-2\sum_p(C^1_{g/p} \otimes f_1^p)(x,Q_i^2) \Bigg)\Bigg\}.\label{inter} 
\end{eqnarray}
 Scale evolution of the collinear PDFs are given by the Dokshitzer-Gribov-Lipatov-Altarelli-Parisi (DGLAP) equation. Using this evolution equation, one can evolve $f_1^g(x,Q_i^2)$ from  the initial scale $Q_i$ to the final  scale $Q_f$ where,  $Q_i<Q_f$
\begin{equation}
f_1^g(x,Q_i^2) = f_1^g(x,Q_f^2)-\frac{\alpha_s}{2\pi} (P_{gg} \otimes f_1^g + P_{gi} \otimes f_1^i)(x,Q_f^2)~\log\frac{Q_f^2}{Q_i^2} + \mathcal{O}(\alpha_s^2).\label{DGLAP}
\end{equation} 
Here, $P_{gg}$ and $P_{gi}$ are leading order splitting functions which are given as,
\begin{eqnarray}
&&P_{gg}(\hat{x})=2C_A\Big[\frac{\hat{x}}{(1-\hat{x})_+}+\frac{1-\hat{x}}{\hat{x}}+\hat{x}(1-\hat{x})\Big]+\delta(1-\hat{x})\frac{11C_A-4n_fT_R}{6}, \label{PGG}\\
&&P_{gq}(\hat{x})=P_{g\bar{q}}(\hat{x})=C_F\frac{1+(1-\hat{x})^2}{\hat{x}}\ . \label{PGQ}
\end{eqnarray}
Where, $C_F=(N_c^2-1)/2N_c$ and $T_R = 1/2$. 
In Eq. (\ref{PGG}), the first term involves the plus prescription and thus avoids an infrared divergence  because of $(1-\hat{x})$  in denominator. 
The plus prescription is given as \cite{boer2020j},
\begin{eqnarray}
\int_y^1dz\frac{G(z)}{(1-z)_+}=\int_y^1dz\frac{G(z)-G(1)}{1-z} - G(1)\log{\Bigg(\frac{1}{1-z}\Bigg)}.
\end{eqnarray}
The $\otimes$ symbol denotes convolution of the two quantities;
\begin{equation}
(P_{gg} \otimes f_1^g)(x,Q^2)=\int_x^1 \frac{d\hat{x}}{\hat{x}}~ P(\hat{x},Q^2)~f\Big(\frac{x}{\hat{x}},Q^2\Big).
\end{equation}
After convolution and substitution of the Eq. (\ref{DGLAP}) in the Eq. (\ref{inter}), we have the  final equation for the unpolarized gluon TMD  as,
\begin{eqnarray}
\hat{f}_1^g(x,\bm{\mathrm{b}}_t^2;Q_f^2) =&& \frac{1}{2\pi} \Bigg\{f_1^g(x,Q_f^2)-\frac{\alpha_s}{2\pi}\Bigg[\Bigg(\frac{C_A}{2}~\log^2\frac{Q_f^2}{Q_i^2}-\frac{11C_A-2n_f}{6}~\log\frac{Q_f^2}{Q_i^2}\Bigg)f^g_1(x,Q_f^2)\nonumber\\
&&~~~~~~~~~~~~~~~~~+(P_{gg} \otimes f_1^g + P_{gi} \otimes f_1^i)(x,Q_f^2)~\log\frac{Q_f^2}{Q_i^2}-2f_1^g(x,Q_f^2) \Bigg]\Bigg\}. \label{f1g}
\end{eqnarray}
Now,  we can write the above equation in the $\bm{\mathrm{q}}_t$-space by making one-to-one correspondence between the functions in impact parameter and  momentum space using  a general formula \cite{tangerman1995intrinsic,van2016quark,Echevarria:2015uaa} 
\begin{eqnarray}
    \hat{f}^{(n)}(x,\bm{\mathrm{b}}_t^2)\equiv \frac{2\pi n!}{M^{2n}}\int_0^{\infty} d \mathrm{q_t}\mathrm{q_t}\left(\frac{\mathrm{q_t}}{\mathrm{\mathrm{b}_t}}\right)^nJ_n(\mathrm{q_t}\mathrm{\mathrm{b}_t})f(x,\bm{\mathrm{q}}_t^2),
    \label{gen}
\end{eqnarray}
here, $n$  is the rank of function in   $\bm{\mathrm{q_t}}$-space.  Since the unpolarized vector-meson
production generally has a rank-zero structure, we can write the unpolarized gluon TMD together with the non-perturbative Sudakov factor in terms of $\bm{\mathrm{q_t}}$-space as
\begin{eqnarray}
f_1^g(x,\bm{\mathrm{q}}_t^2)=\frac{1}{2\pi}&&\int^\infty_0 \mathrm{b}_td\mathrm{b}_t J_0(\mathrm{b}_t\mathrm{q}_t) \Bigg\{f_1^g(x,Q_f^2)\nonumber\\
&&~~~-\frac{\alpha_s}{2\pi}\Bigg[\Bigg(\frac{C_A}{2}~\log^2\frac{Q_f^2}{Q_i^2}-\frac{11C_A-2n_f}{6}~\log\frac{Q_f^2}{Q_i^2}\Bigg)f^g_1(x,Q_f^2)\nonumber\\
&&~~~~+(P_{gg} \otimes f_1^g + P_{gi} \otimes f_1^i)(x,Q_f^2)~\log\frac{Q_f^2}{Q_i^2}
-2f_1^g(x,Q_f^2) \Bigg]\Bigg\}\times e^{-S_{np}(\bm{\mathrm{b}}_t^2)}.\label{finalf1g}
\end{eqnarray}
Now let us write the expression for the linearly polarized gluon distribution function $h_1^{\perp g}(x,\bm{\mathrm{b}}_t^2)$. 
The perturbative tail of $h_1^{\perp g}$ can be computed in the same way as the perturbative tail of $f_1^{ g}$, with the key difference  that its expansion in powers of the QCD coupling constant begins at $\mathrm{O}(\alpha_s)$. Using Eq. (3.13) of Ref. \cite{boer2020j} and then performing the Fourier transformation using  Eq. (\ref{gen}), we write the linearly polarized gluon distribution at LO in terms of the unpolarized
collinear PDFs $f_1^a(\hat{x},Q_f^2)$ in the $\bm{\mathrm{b}}_t$-space as,

\begin{eqnarray}
h_1^{\perp g(2)}(x,\bm{\mathrm{b}}_t^2;Q_f^2)=&&\frac{2\alpha_S}{\pi^2M_p^2}\frac{1}{\bm{\mathrm{b}}_t^2}\Bigg[C_A\int_x^1\frac{d\hat{x}}{\hat{x}}\Bigg(\frac{\hat{x}}{x}-1\Bigg)f_1^g(\hat{x},Q_f^2)\nonumber\\
&&~~~~~~~~~~~~~~~~~~~~+C_F\sum_{p=q,\bar{q}}\int_x^1\frac{d\hat{x}}{\hat{x}}\Bigg(\frac{\hat{x}}{x}-1\Bigg)f_1^p(\hat{x},Q_f^2)\Bigg].\label{h1perpbt}
\end{eqnarray}
Here we have used the general formula for Bessel integral \cite{kovchegov2012quantum}
\begin{eqnarray}
    \int_0^{\infty} dk~k^{\lambda-1} ~J_{\nu}(kx)=2^{\lambda-1} x^{-\lambda}\frac{\Gamma(1/2(\nu+\lambda))}{\Gamma(1/2(2+\nu+\lambda))},
\end{eqnarray}
where $J_{\nu}(z)$ is the Bessel function of the first kind of order $\nu$. The $\mathrm{q}_t$ dependence for $h_1^{\perp g}$
in the gluon correlator has a rank-two
tensor structure in the non-contracted transverse momentum, thus
  we could write the linearly polarized gluon TMD  $h_1^{\perp g}$ in the $\bm{\mathrm{q}}_t$-space is given as,
\begin{eqnarray}
    \frac{\bm{\mathrm{q}}_t^2}{M_p^2}h_1^{\perp g(2)}(x,\bm{\mathrm{q}}_t^2)=&& \frac{\alpha_s}{\pi^2} 
    \int_0^{\infty} d {\mathrm{b_t}}{\mathrm{b_t}}~J_2(\mathrm{q_t}\mathrm{b_t})~\Bigg[C_A\int_x^1\frac{d\hat{x}}{\hat{x}}\Bigg(\frac{\hat{x}}{x}-1\Bigg)f_1^g(\hat{x},Q_f^2)\nonumber\\ &&~~~~~~~~~~~~~~~~~~~~+C_F\sum_{p=q,\bar{q}}\int_x^1\frac{d\hat{x}}{\hat{x}}\Bigg(\frac{\hat{x}}{x}-1\Bigg)f_1^p(\hat{x},Q_f^2)\Bigg]\times e^{-S_{np}(\bm{\mathrm{b_t}}^2)}.\label{finalh1perpg}
\end{eqnarray}
\subsection{Non-perturbative Sudakov factor}

In the above Eqs.~\ref{finalf1g} and \ref{finalh1perpg}, the perturbative part is strictly valid in the perturbative domain, which means low $\mathrm{b}_t$ (or $\mathrm{b}_t<<1/\Lambda_{QCD}$). However, in order to perform the corresponding Fourier transform, we need to integrate  the expression from small to large $\mathrm{b}_t$. As a result, the perturbative expression for the Sudakov factor given above should not be used alone rather one needs to introduce a non-perturbative Sudakov factor, this should suppress the large $\mathrm{b}_t$ domain. 
The functional form of the non-perturbative Sudakov factor is constrained by two conditions, one of which is it has to be equal to $1$ for $\mathrm{b}_t = 0$ and for large $\mathrm{b}_t$ it is supposed to decrease monotonically and ultimately should vanish. The functional form attributed to $S_{np}$ is quadratic in $\mathrm{b}_t$ with $e^{-S_{np}}$ reaching 0 within a certain value of $\mathrm{b}_t$ called $\mathrm{b}_{t\text{lim}}$. 
However, when $\mathrm{b}_t$ gets too small, the lower scale $Q_i=2e^{-\gamma_E}/\mathrm{b}_t$ will be larger than final scale $Q_f$ and hence, we expect the evolution should stop. This could be resolved by taking a $\mathrm{b}_t$ prescription as given below \cite{scarpa2020studies},
\begin{eqnarray}
b_{t\star}(\mathrm{b}_c(\mathrm{b}_t))=\frac{\mathrm{b}_c(\mathrm{b}_t)}{\sqrt{1+\frac{b_c(\mathrm{b}_t)}{b_{t\text{max}}}}}
\end{eqnarray}
where,
\begin{eqnarray}
&&\mathrm{b}_c=\sqrt{\mathrm{b}_t^2+\Big(\frac{2e^{-\gamma_E}}{Q_f}\Big)^2}.\nonumber
\end{eqnarray}
This prescription constrains the $Q_i(\mathrm{b}_t)=2\gamma_E/b_{t\star}(\mathrm{b}_t)$ range in between $2\gamma_E/\mathrm{b}_{t\text{max}}$ (for $\mathrm{b}_t\to \infty$) and $Q_f$ (for $\mathrm{b}_t\to 0$). 
Motivated by \cite{scarpa2020studies} we choose the Gaussian behavior of $e^{-S_{np}}$  as ,
\begin{eqnarray}
&&S_{np}=\frac{A}{2}\log{\Big(\frac{Q_f}{Q_{np}}\Big)}\mathrm{b}_c^2,~~~~~~~~~\quad Q_{np}=1~\mathrm{GeV}. \label{SNP}
\end{eqnarray}
The parameter $A$ controls the width of the non-perturbative Sudakov factor for a particular $Q_f$. In this calculation, we have taken the final scale as $Q_f =  \sqrt{M_\psi^2+\mathrm{K}_{t}^2}$. We have taken $A=2.3$ GeV$^2$, which is calculated for $\mathrm{K}_t = 1~\mathrm{GeV} $. The $\mathrm{K}_t$ dependence present in $Q_f$ does not affect the value of $A$, as $\mathrm{K}_t$ increases the $e^{-S_{np}}$ remains below the convergence criteria at the given $\mathrm{b}_{t\text{lim}}$. 
The value of $\mathrm{b}_{t\text{max}}$ for the following numerical study is $1.5~\mathrm{GeV}^{-1}$ which is in consistent with \cite{scarpa2020studies}, Collins-Soper-Sterman formalism given for Z Boson production \cite{Collins-Soper-Sterman2003} and Collins-Soper-Sterman formalism implemented by \cite{Aybat2011}. Below we present two recent parameterizations of the gluon TMDs that we have used. 

		\section{Spectator model}
		In this section, we discuss a recent parameterization of the gluon TMDs based on a spectator model \cite{Bacchetta:2020vty}.
		According to this model, the remnant after the gluon emission from the nucleon is treated as a single spectator particle, which is on-shell, with mass $M_X$.
		The mass can take a range of values given by a spectral function.
		The nucleon-gluon-spectator coupling is encoded in an effective vertex that contains two form factors.
		The expression for a given TMD reads as
		\begin{eqnarray}
			F^g (x, \bm{\mathrm{q}}_t^2) = \int_M^\infty dM_X \, \rho_X ( M_X ) \, \hat{F}^g (x, \bm{\mathrm{q}}_t^2; M_X) \; .
			\label{eqweightedTMD}
		\end{eqnarray}
		Here $\rho_X ( M_X )$ is the spectral function and can be written as 
		\begin{eqnarray}
			\rho_X (M_X) =  \mu^{2a} \left[ \frac{A}{B + \mu^{2b}} + \frac{C}{\pi \sigma} e^{- \frac{(M_X - D)^2}{\sigma^2}} \right] \;.
			\label{eqspectral}
		\end{eqnarray}
		Where $\mu=M_X^2-M^2$ and $\{X\}\equiv \{A, B, a, b, C, D, \sigma \}$ are
free parameters. The parameter B in the above equation is set at $B = 2.1$ and $M_X$ can take real values in the continuous range according to the above spectral function. The nucleon mass M is taken to be 1.  The parameters a, b have a strong influence on the spectral function, at larger $M_X$, its asymptotic trend depends on the sign of the  difference $a-b$ \cite{Bacchetta:2020vty}. For $a-b<0$, the value of $\rho_X$ approaches zero for large $M_X$. We consider integration in Eq.(\ref{eqweightedTMD}) over the range $1<M_X<10$ GeV. The value of different parameters that we used for our numerical results are given in the Table \ref{tabpar-g} below. These model parameters have been fixed by fitting the NNPDF data at scale $Q=1.64$ GeV \cite{Bacchetta:2020vty}. We assumed the same set of parameters to probe the TMDs at $Q=\sqrt{M_\psi^2+\mathrm{K}_t^2}$. In this model, we don't have direct inference of scale dependency on the TMDs unlike the case of Gaussian parameterization. However, the longitudinal momentum fraction $x$ depends on $Q$ but going from some low $Q=1.64$ GeV to some relatively large $Q=\sqrt{M_\psi+\mathrm{K}_t^2}$ ($\approx 6.75$ GeV for $M_\psi=3.1$ GeV and $\mathrm{K}_t=6$ GeV) hardly changes the $x$ range.  
		\begin{table}[H]
			\centering
			\begin{tabular}[c]{c|c|c|c}
			\toprule
				\hline \hline
				Parameter      & Replica 11 & Parameter & Replica 11\\
				\hline\hline
				$A$  &  6.0 &$\kappa_2$ (GeV$^2$)& 0.414\\
				$a$ & 0.78 &$\sigma$ (GeV) & 0.50 \\
				$b$ & 1.38&$\Lambda_X$ (GeV) & 0.448 \\
				$C$& 346&$\kappa_1$ (GeV$^2$)& 1.46  \\
				$D$ (GeV)  & 0.548 && \\    	
				\hline \hline
			\end{tabular}
			\caption{Corresponding values for replica 11.}
			\label{tabpar-g}
		\end{table}
		The leading-twist T -even unpolarized and linearly polarized gluon TMDs can be written as\cite{mulders2001transverse,meissner2007relations}
		\begin{eqnarray}
			\hat{f}_1^g (x, \bm{\mathrm{q}}_t^2; M_X) &=&- \frac{1}{2}\, g^{i j} \, \left[ \Phi^{i j} (x, \bm{\mathrm{q}}_t, S) + \Phi^{i j} (x, \bm{\mathrm{q}}_t, -S) \right]  \nonumber \\
			&=&\Big[ \big( 2 M x g_1 - x (M+M_X) g_2 \big)^2 \, \big[ (M_X - M (1-x) )^2 + \bm{\mathrm{q}}_t^2 \big]\nonumber\\
			&&~~~~~~~~~~~ +2 \bm{\mathrm{q}}_t^2 \, (\bm{\mathrm{q}}_t^2 + x M_X^2) \, g_2^2 + 2 \bm{\mathrm{q}}_t^2 M^2 \, (1-x) \, (4 g_1^2 - x g_2^2 ) \Big]  \nonumber \\ 
			&&~~~~~~~~~~~~~~~~~~~~~\times \Big[ (2 \pi)^3 \, 4 x M^2 \, (L_X^2 (0) + \bm{\mathrm{q}}_t^2)^2 \Big]^{-1},
			\label{eqf1}
		\end{eqnarray}
		\begin{eqnarray}
			\hat{h}_1^{\perp g} (x, \bm{\mathrm{q}}_t^2; M_X) &=&\frac{M^2}{\varepsilon_t^{i j} \delta^{j m} (p_t^j p_t^m + g^{j m} \bm{\mathrm{q}}_t^2)} \, \varepsilon_t^{l n} \delta^{n r} \, \left[ \Phi^{n r} (x, \bm{\mathrm{q}}_t, S) + \Phi^{n r} (x, \bm{\mathrm{q}}_t, -S) \right] \nonumber \\
			&=&\Big[ 4 M^2 \, (1-x) \, g_1^2 + ( L_X^2 (0) + \bm{\mathrm{q}}_t^2 ) \, g_2^2 \Big]\times \Big[ (2 \pi)^3 \, x \, (L_X^2 (0) + \bm{\mathrm{q}}_t^2)^2 \Big]^{-1}.
			\label{eqh1perp}
		\end{eqnarray}
		Here  $g_{1,2} (p^2)$ are model-dependent form factors and can be written as 
		\begin{equation}
			g_{1,2} (p^2) = \kappa_{1,2} \, \frac{p^2}{|p^2 - \Lambda_X^2|^2} = \kappa_{1,2} \, \frac{p^2 \, (1 - x)^2}{(\bm{\mathrm{q}}_t^2 + L_X^2 (\Lambda_X^2))^2}  \; ,
			\label{eqdipolarff}
		\end{equation}
		where, $\kappa_{1,2}$ and $\Lambda_X$ are normalization and cut-off parameters, respectively, and 
		\begin{eqnarray}
		p^2&&=-\frac{\bm{\mathrm{q}}^2_t+L_X^2 (0)}{1-x},
		\end{eqnarray}
		where p is the gluon momentum and
		\begin{eqnarray}
			L_X^2 (\Lambda_X^2)&& = x \, M_X^2 + (1 - x) \, \Lambda_X^2 - x \, (1 - x) \, M^2 \;.
			\label{eqLX}
			\end{eqnarray}
			The form factors given above are dipolar in nature;  
		the main advantage of using dipolar form factors
		consists in the possibility of canceling gluon-propagator singularities, quenching the effects of
		large transverse momenta where a pure TMD description is not any more adequate, and removing
		logarithmic divergences emerging in $p_t$ -integrated densities.
		
		\section{Gaussian parameterization of the TMDs}
		The most widely used parameterization of the TMDs are Gaussian in nature, Here,  both the TMDs, $f_1^g$ and $h_1^{\perp g}$ are assumed to be  factorized into a product of a $x$ dependent part given in terms of the collinear pdfs and an  exponential factor which is a function of only the transverse momentum  $(\bm{\mathrm{q}}_t)$. The width of the Gaussian is usually expressed in terms of the average value of the transverse momentum, which is taken as a model  parameter  \cite{boer2012polarized,mukherjee2017linearly,mukherjee2016probing}
		\begin{eqnarray}
			\label{g1}
			f_1^g(x,\textbf{q}_{t}^2)=f_1^g(x,\mu)\frac{1}{\pi\langle \bm{\mathrm{q}}_t^2\rangle}e^{-\bm{\mathrm{q}}_t^2/\langle \bm{\mathrm{q}}_t^2\rangle},
		\end{eqnarray}
		\begin{eqnarray}
			\label{g2}
			h_1^{\perp g}(x,\bm{\mathrm{q}}_{t}^2)=\frac{M_p^2f_1^g(x,\mu)}{\pi\langle \bm{\mathrm{q}}_t^2\rangle^2}\frac{2(1-r)}{r}e^{1-\frac{\bm{\mathrm{q}}_t^2}{r\langle \bm{\mathrm{q}}_t^2\rangle}},
		\end{eqnarray}
		where, $r(0<r<1)$ is a parameter and in our case we take $r=1/3$. The term  $f_1^g(x,\mu)$ is the collinear PDF which follows the DGLAP evolution  equation. 
		The Gaussian width here is  $\langle \bm{\mathrm{q}}_t^2 \rangle=0.25 ~ \mathrm{GeV}^2$. The linearly-polarized gluon distribution in the
parameterization above satisfies the positivity bound \cite{mulders2001transverse}, but
does not saturate it.
\begin{eqnarray}\label{eq:ub}
\frac{\bm{\mathrm{q}}_{t}^2}{2M_p^2}|h_1^{\perp g}(x,\bm{\mathrm{q}}_{t}^2)|\leq 	f_1^g(x,\textbf{q}_{t}^2).
\end{eqnarray}
\section{Results and Discussion}
In the present work, we have numerically calculated the  $\cos2\phi_t$ azimuthal asymmetry in the unpolarized electroproduction of $J/\psi$ process: $e~p\rightarrow e~J/\psi~Jet~ X$, within TMD factorization approach. As depicted in  Fig. \ref{figbtb}, we have $J/\psi$ and jet almost back-to-back in the transverse plane as  we consider the kinematics,  $|\bm{\mathrm{q}}_t|\ll |\bm{\mathrm{K}}_t|$ which is the required condition to assume TMD factorization.  Contributions from the virtual photon-quark(anti-quark) initiated sub-processes in the unpolarized cross-section is very small in the kinematics considered \cite {DAlesio:2019qpk} as compared with gluon initiated subprocess: $\gamma^* + g  \rightarrow J/\psi + g$. Therefore, in the numerical estimate of the asymmetry, we have included only the gluon-photon fusion subprocess and neglected the contribution from the quark(anti-quark) initiated sub-processes. 
We have used MSTW2008\cite{martin2009parton} set of collinear PDFs and adopted two sets of LDMEs for the study of azimuthal asymmetry as listed in Table \ref{tabLDME} with the charm mass $m_c = 1.3$ GeV. We used NRQCD framework  for $J/\psi$ production rate and included contributions from both color singlet and color octet states in the asymmetry. We also calculated the asymmetry taking into consideration  contribution only from the color singlet state (CS); and compared it with  the full NRQCD result incorporating both CS and CO contributions (NRQCD). The contraction of the different states, i.e., $^1S_0^{(8)},~^3 S_1^{(1,8)}$ and $^3P^{(8)}_{J(=0,1,2)}$ is calculated using  FeynCalc\cite{shtabovenko2020feyncalc,mertig1991feyn}. We have investigated the effect of TMD evolution on the asymmetry. For the gluon TMDs we use two parameterizations, Gaussian and based on the spectator model, as discussed above. The $J/\psi$ mass is taken to be $M_{\psi}=3.1$ GeV. The collinear PDFs are evaluated at the scale $Q=\sqrt{M_{\psi}^2+\mathrm{K}_t^2}$. The numerical results are presented in the kinematical region that can be accessed at the future EIC. 

We have imposed a cut on the variable $z$, namely,  $0.1<z<0.9$, to estimate the asymmetry. As $z\to 1$, the final state gluon becomes soft, which leads to infrared divergences. We impose the upper cut to avoid this gluon to become soft. Contribution of $J/\psi$ production from fragmentation of final hard gluon comes from lower $z$ region, and we impose the lower cut to minimize this contribution. The asymmetry gets maximized around $z=0.7$ for the kinematics we have considered. Hence, we took $z=0.7$ for all plots where $z$ is fixed. In addition, we also show the $\cos2\phi_t$ asymmetry as function of $z$. In our estimate, we have neglected the contribution of $J/\psi$ production via feed-down from excited $\psi(2S)$ and the decays of $\chi_c$ states.

\begin{figure}[H]
			\begin{minipage}[c]{0.99\textwidth}
				\small{(a)}\includegraphics[width=8cm,height=6.5cm,clip]{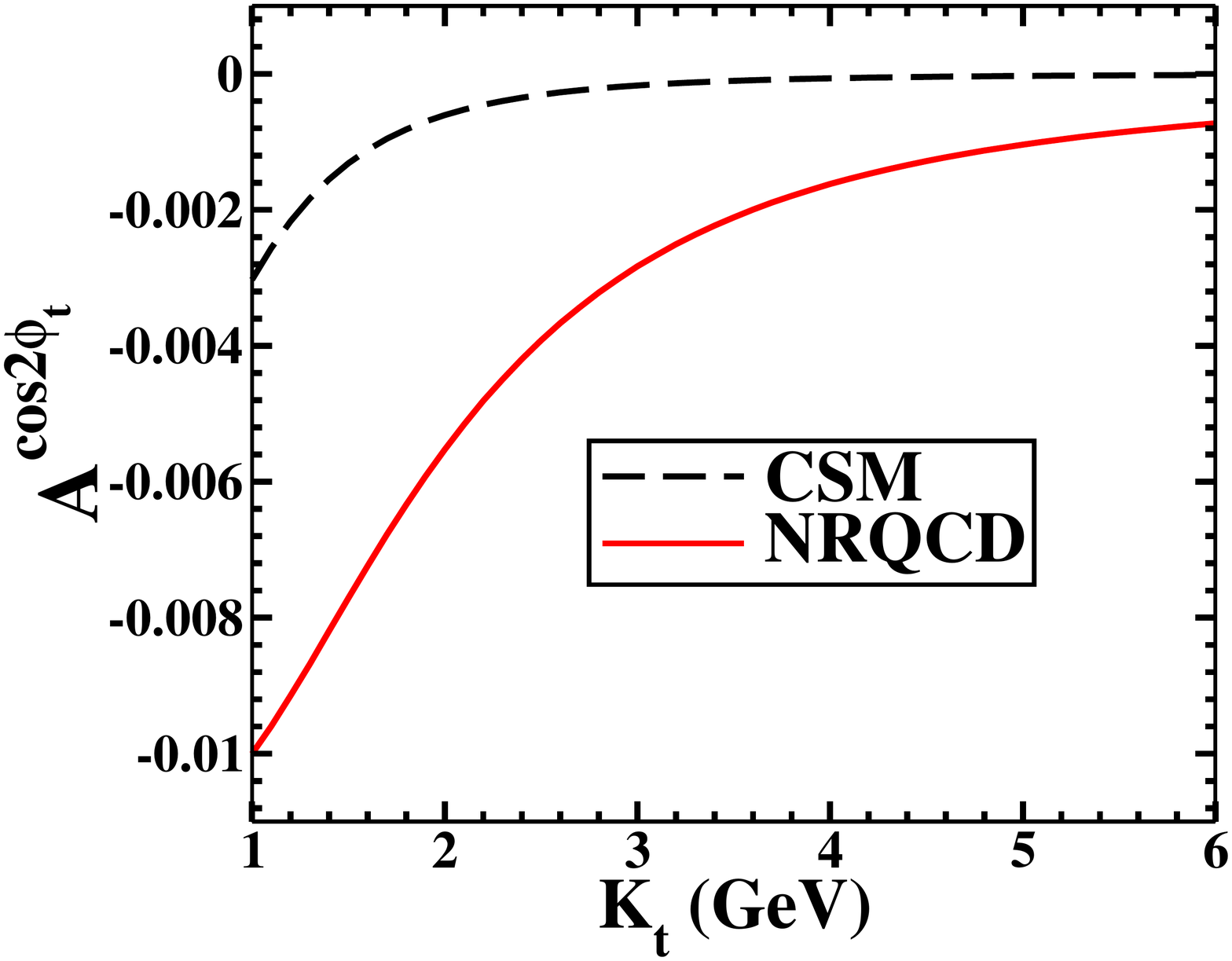}
				\hspace{0.1cm}
				\small{(b)}\includegraphics[width=8cm,height=6.5cm,clip]{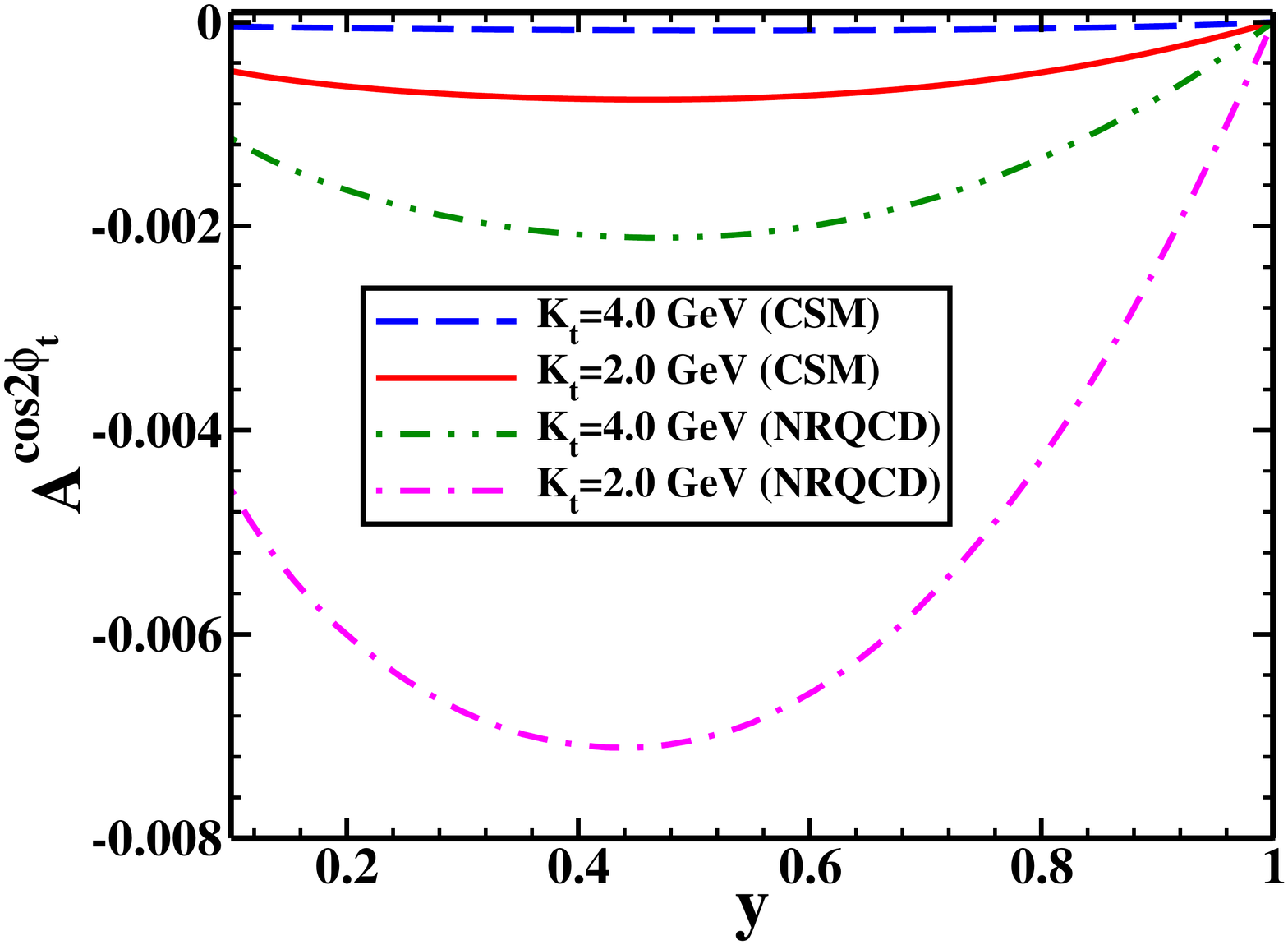}
			\end{minipage}
			\begin{minipage}[c]{0.99\textwidth}
			\centering
				\small{(c)}\includegraphics[width=8cm,height=6.5cm,clip]{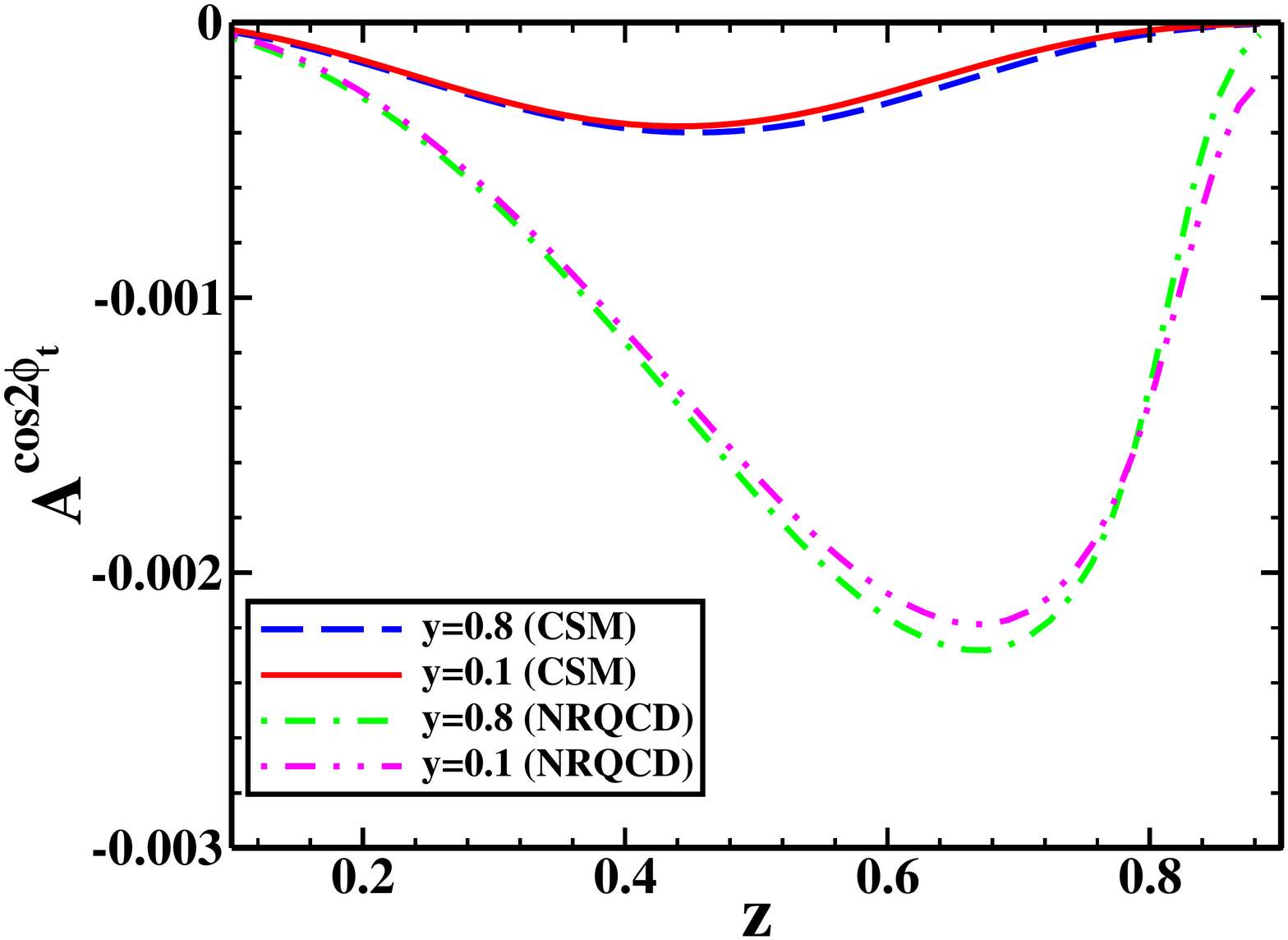}
			\end{minipage}
			\caption{$\cos2\phi_t$ asymmetry calculated in TMD evolution approach for $e+p\rightarrow e+ J/\psi+ Jet+X$, in both NRQCD and in the CS; as functions of (a) $\mathrm{K}_{t}$, (b) $y$ and (c) $z$. We have used $\sqrt{s}=140~$GeV. In (a) and (b) we have used $z=0.7$. In (a) we have taken $0.1\leq y\leq 1$ for the range of $y$ integration and in (b) we have used fixed values of  $\mathrm{K}_t$. In (c) we have taken $\mathrm{K}_t=3~$GeV and fixed values of $y$. We have used CMSWZ set of LDMEs \cite{ChaoLDME}.}
			\label{evol}
		\end{figure}
\begin{figure}[H]
			\begin{minipage}[c]{0.99\textwidth}
				\small{(a)}\includegraphics[width=8cm,height=6.5cm,clip]{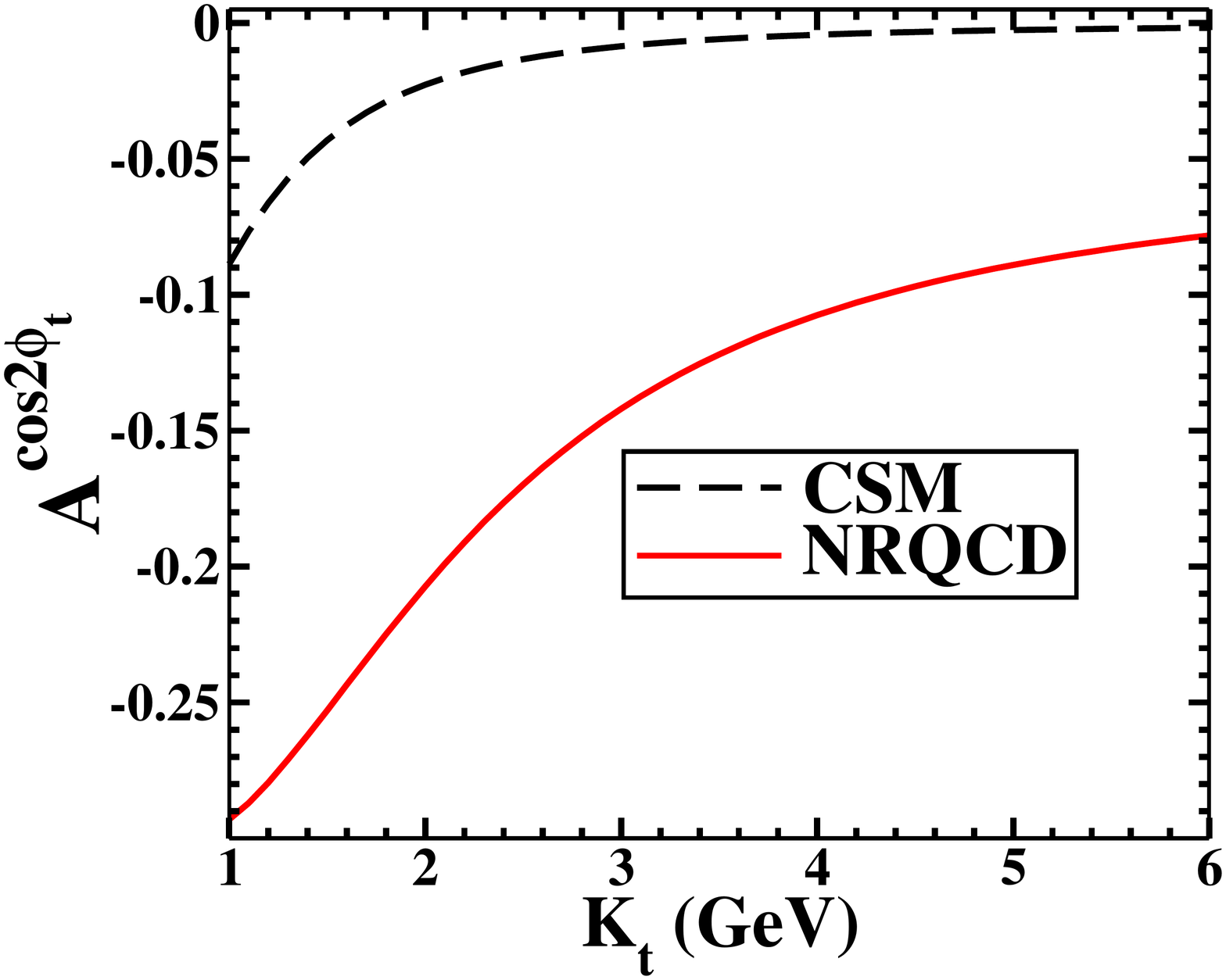}
				\hspace{0.1cm}
				\small{(b)}\includegraphics[width=8cm,height=6.5cm,clip]{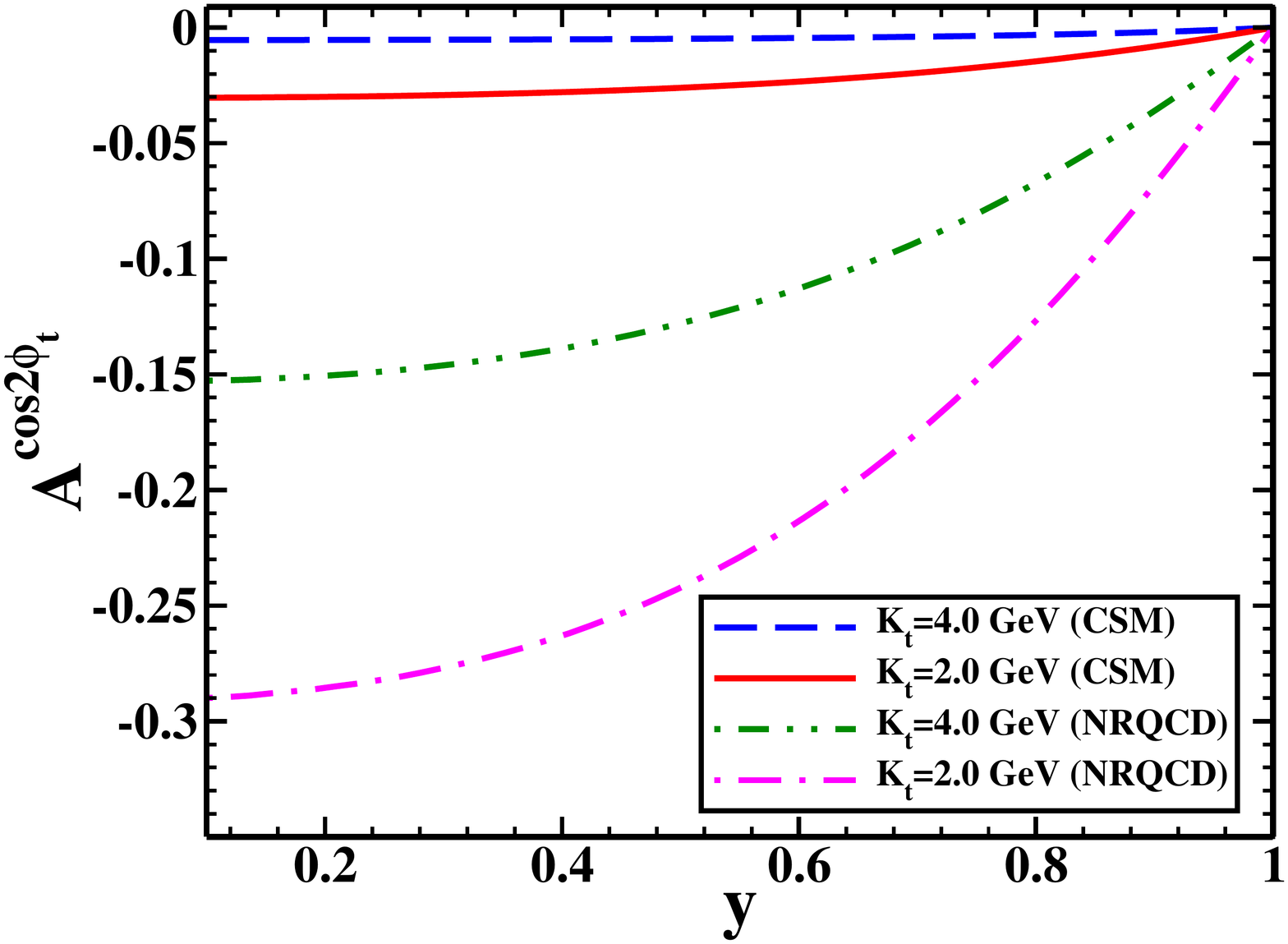}
			\end{minipage}
			\begin{minipage}[c]{0.99\textwidth}
			\centering
				\small{(c)}\includegraphics[width=8cm,height=6.5cm,clip]{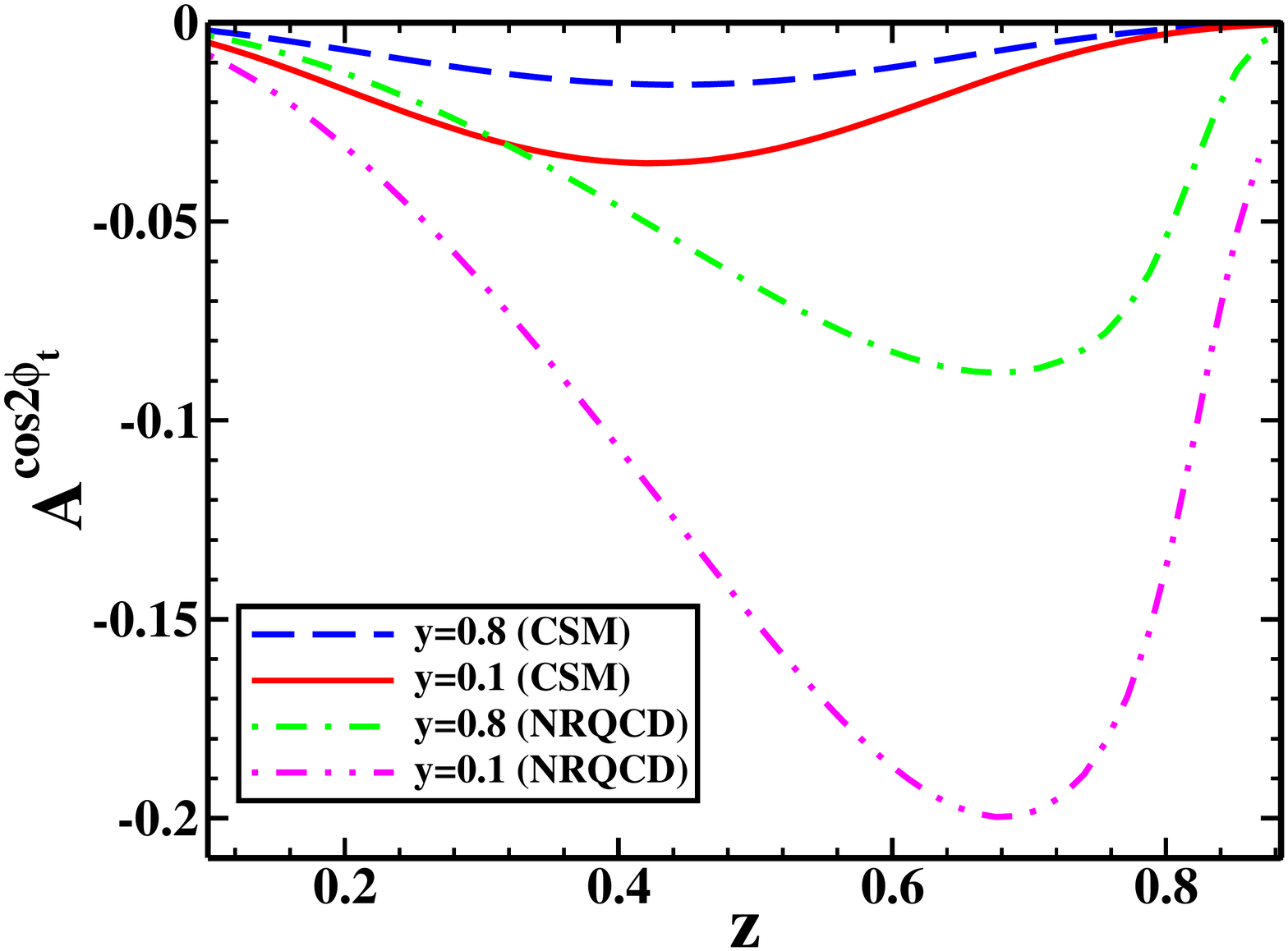}
			\end{minipage}
			\caption{$\cos2\phi_t$ asymmetry calculated in the spectator model for $e+p\rightarrow e+ J/\psi+ Jet+X$ process, in both NRQCD and in the CS; as functions of (a) $\mathrm{K}_{t}$, (b) $y$ and (c) $z$. We have used $\sqrt{s}=140$ GeV. In (a) and (b) we have used $z=0.7$. In (a) we have taken $0.1\leq y\leq 1$ for the range of $y$ integration and in (b) we have used fixed values of $\mathrm{K}_t$. In (c) we have taken $\mathrm{K}_t=3~$GeV and fixed values of $y$. We have used CMSWZ set of LDMEs \cite{ChaoLDME}.}  
			\label{spec}
		\end{figure}
\begin{figure}[H]
			\begin{minipage}[c]{0.99\textwidth}
				\small{(a)}\includegraphics[width=8cm,height=6.5cm,clip]{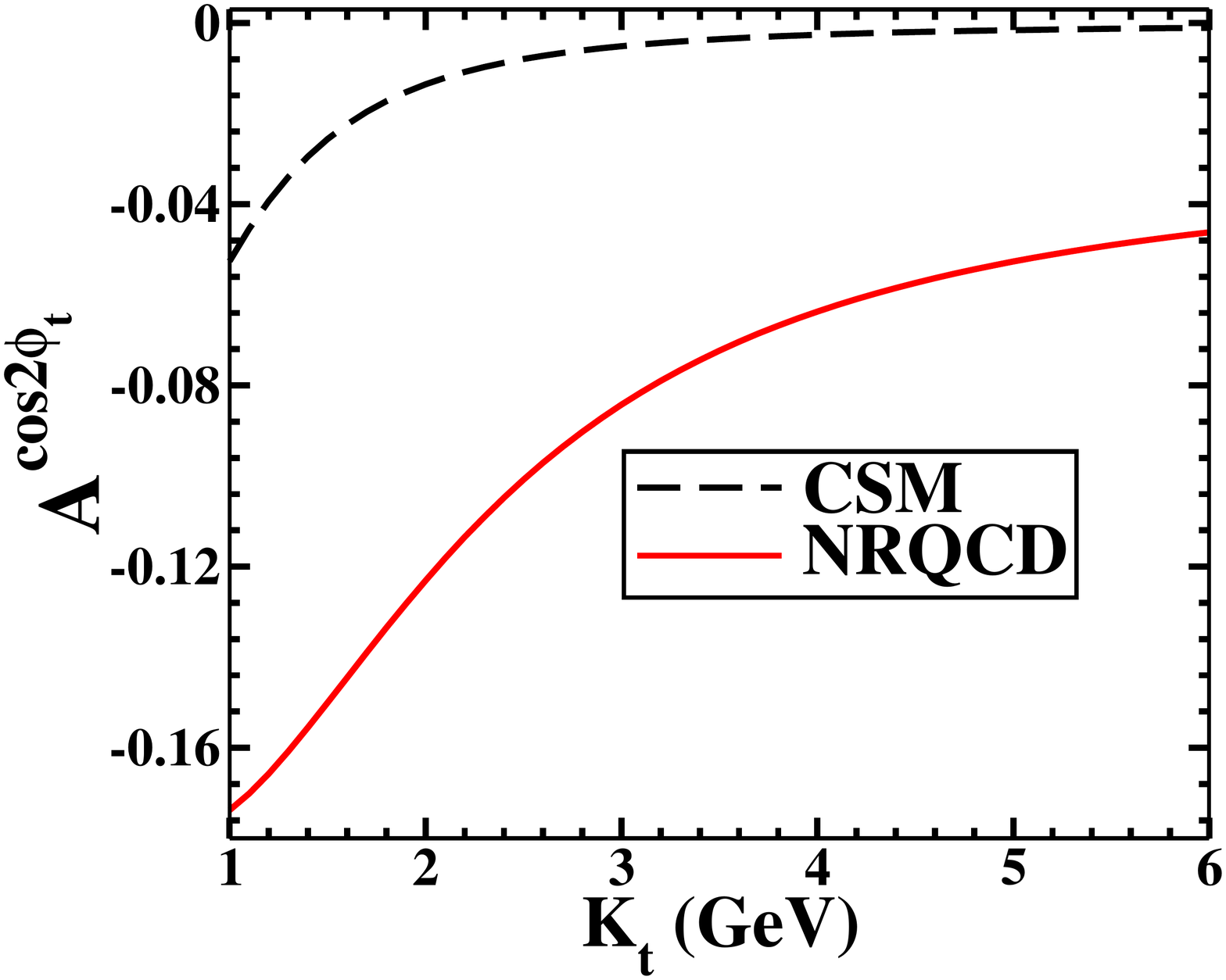}
				\hspace{0.1cm}
				\small{(b)}\includegraphics[width=8cm,height=6.5cm,clip]{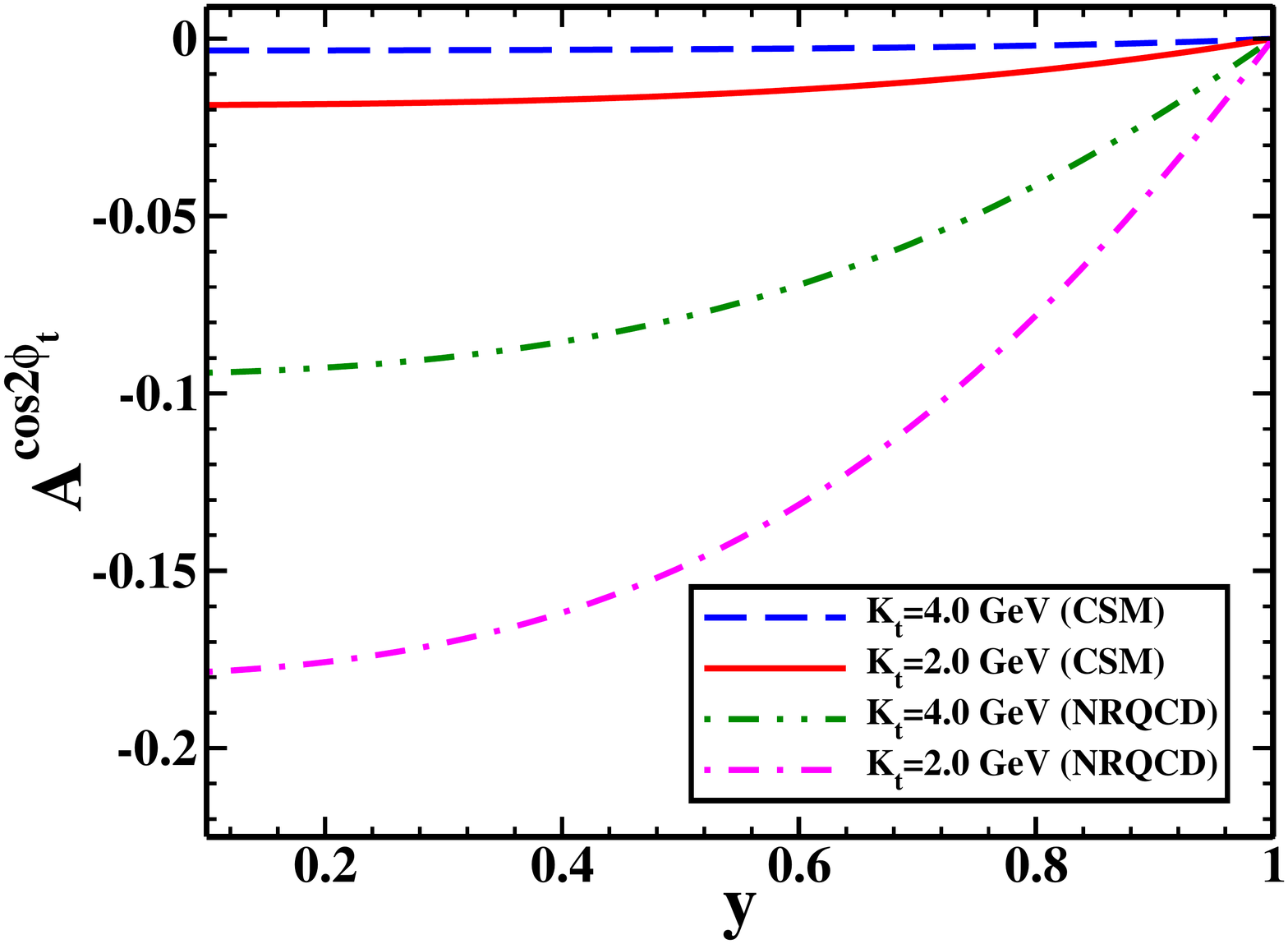}
			\end{minipage}
			\begin{minipage}[c]{0.99\textwidth}
			\centering
				\small{(c)}\includegraphics[width=8cm,height=6.5cm,clip]{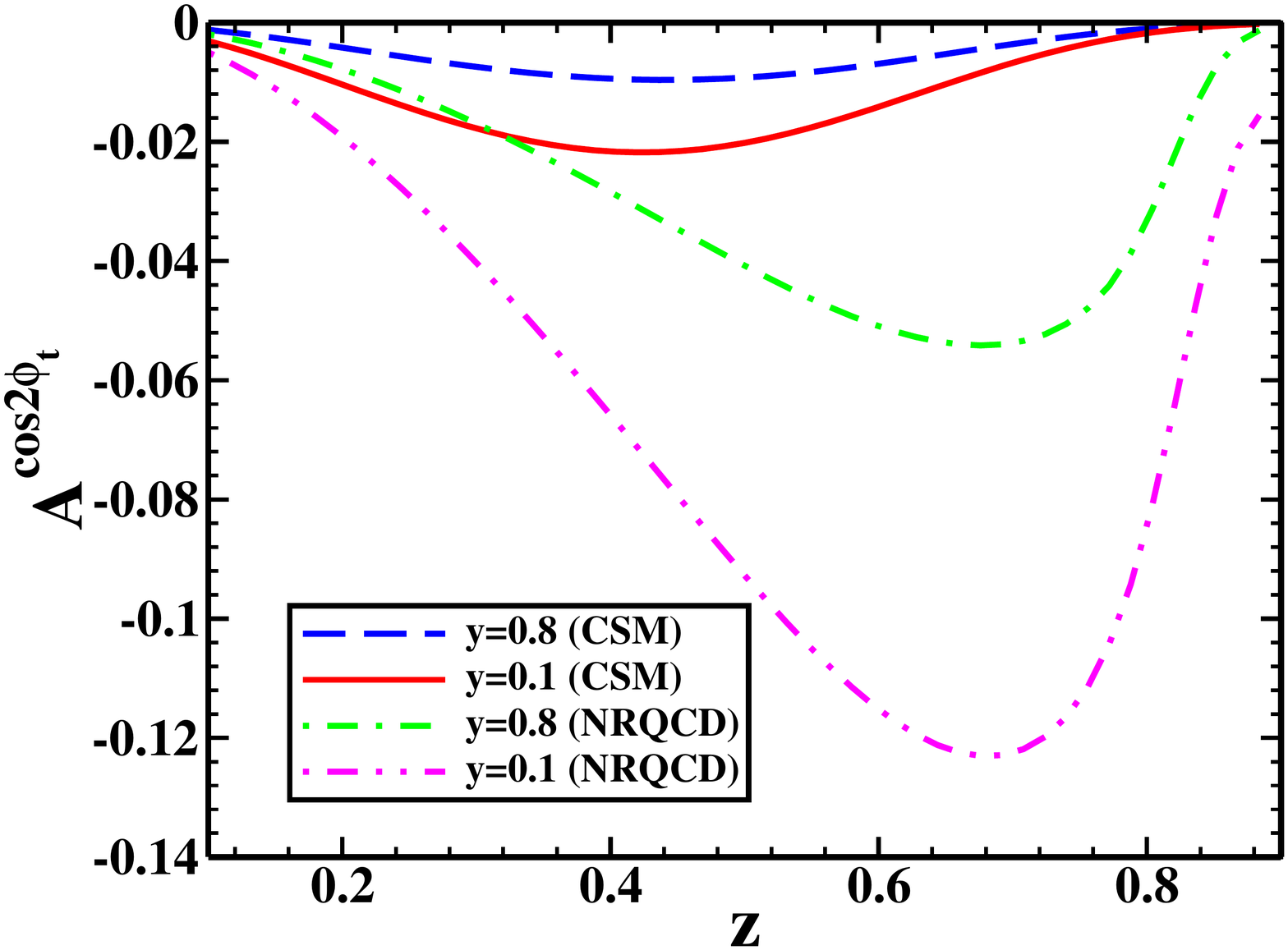}
			\end{minipage}
			\caption{$\cos2\phi_t$ asymmetry calculated using Gaussian  parameterization of TMDs for $e+p\rightarrow e+ J/\psi+ Jet+X$ process, as functions of (a) $\mathrm{K}_{t}$, (b) $y$ and (c) $z$. We have used $\sqrt{s}=140~$GeV. In (a) and (b) we have used $z=0.7$. In (a) we have taken $0.1\leq y\leq 1$ for the range of $y$ integration and in (b) we have used fixed values of $\mathrm{K}_t$. In (c) we have taken $\mathrm{K}_t=3~$GeV and fixed values of $y$. We have used CMSWZ set of LDMEs \cite{ChaoLDME}.}  
			\label{gauss}
		\end{figure}
\begin{figure}[H]
			\begin{minipage}[c]{0.99\textwidth}
				\small{(a)}\includegraphics[width=8cm,height=6.5cm,clip]{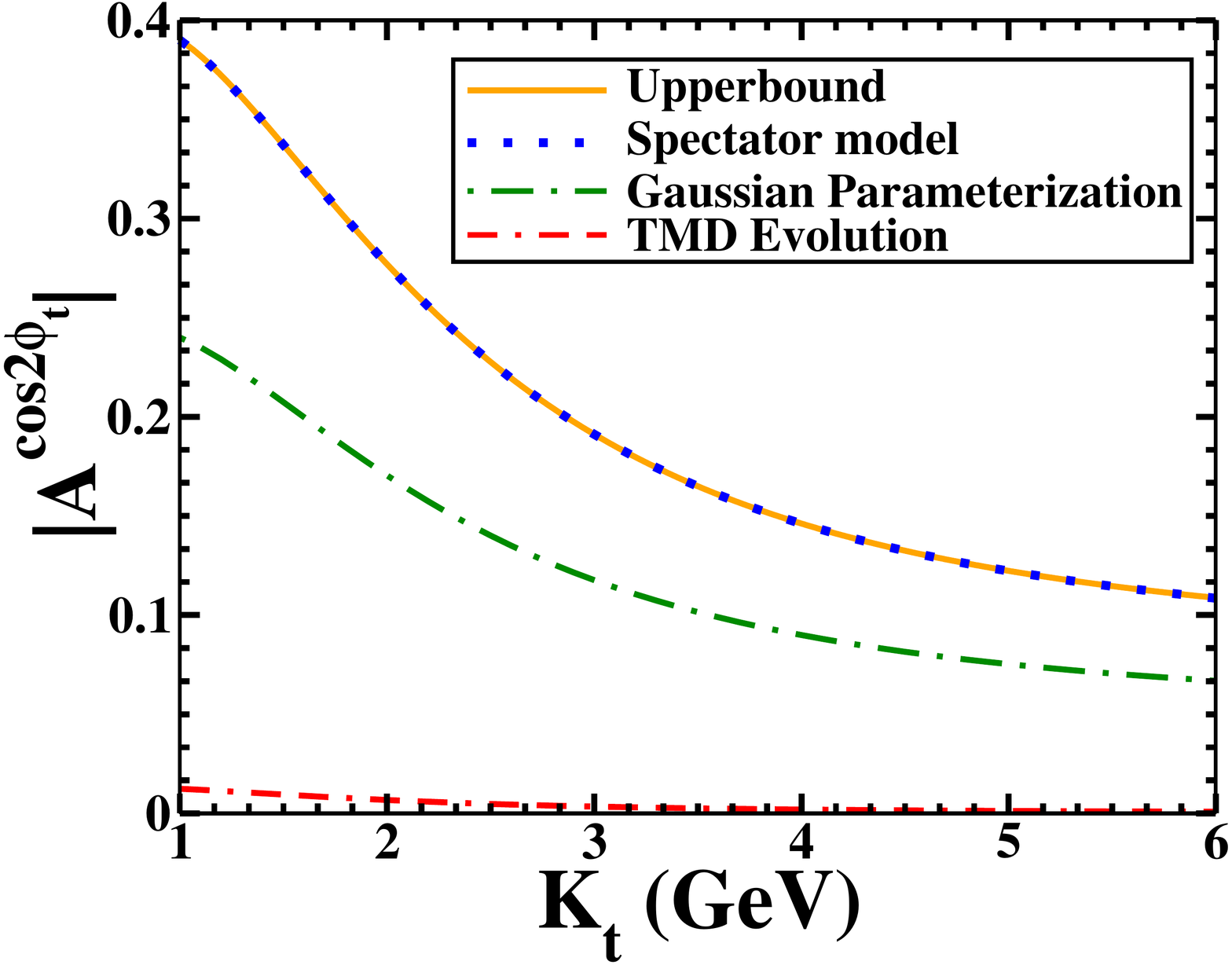}
				\hspace{0.1cm}
				\small{(b)}\includegraphics[width=8cm,height=6.5cm,clip]{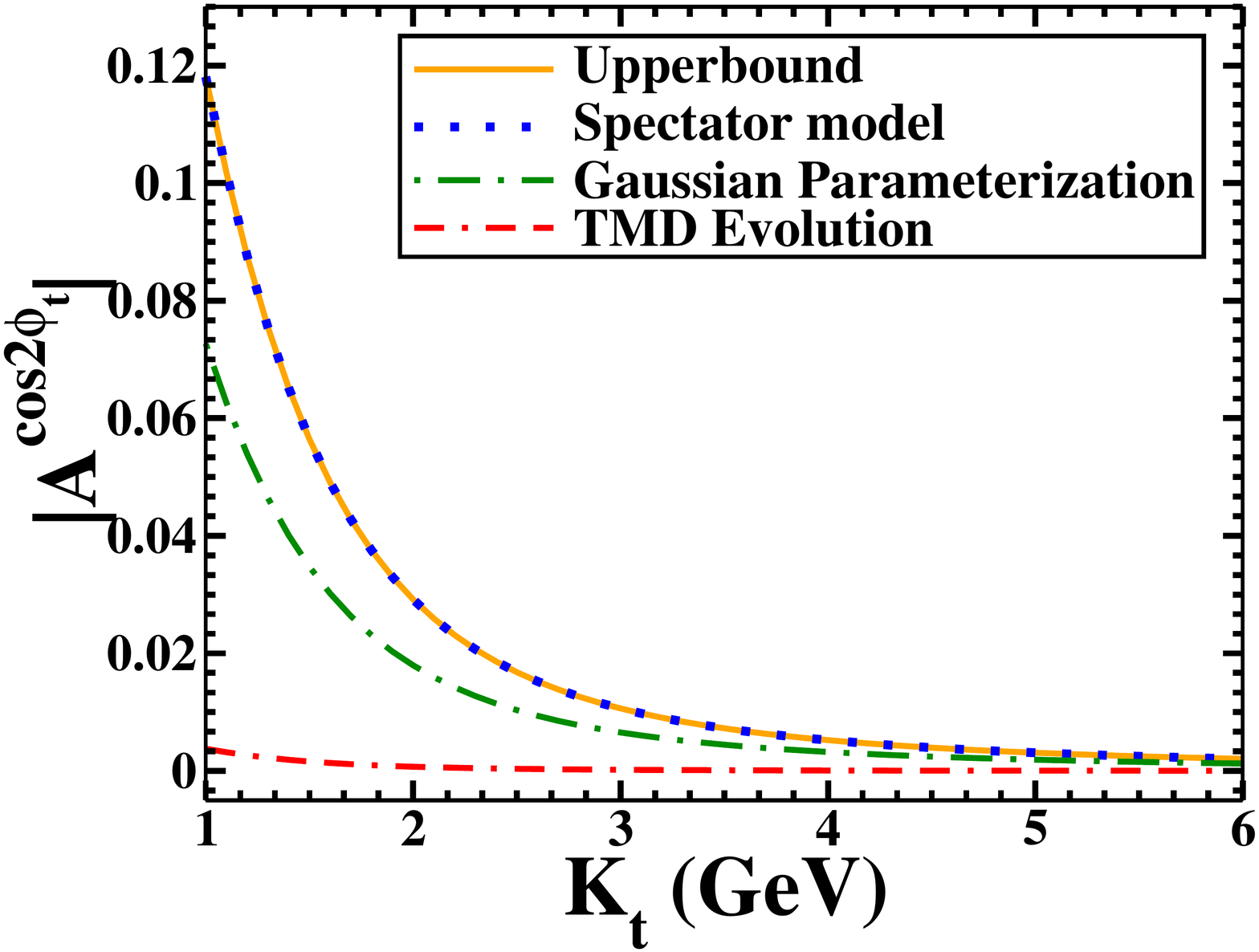}
			\end{minipage}
			\begin{minipage}[c]{0.99\textwidth}
				\small{(c)}\includegraphics[width=8cm,height=6.5cm,clip]{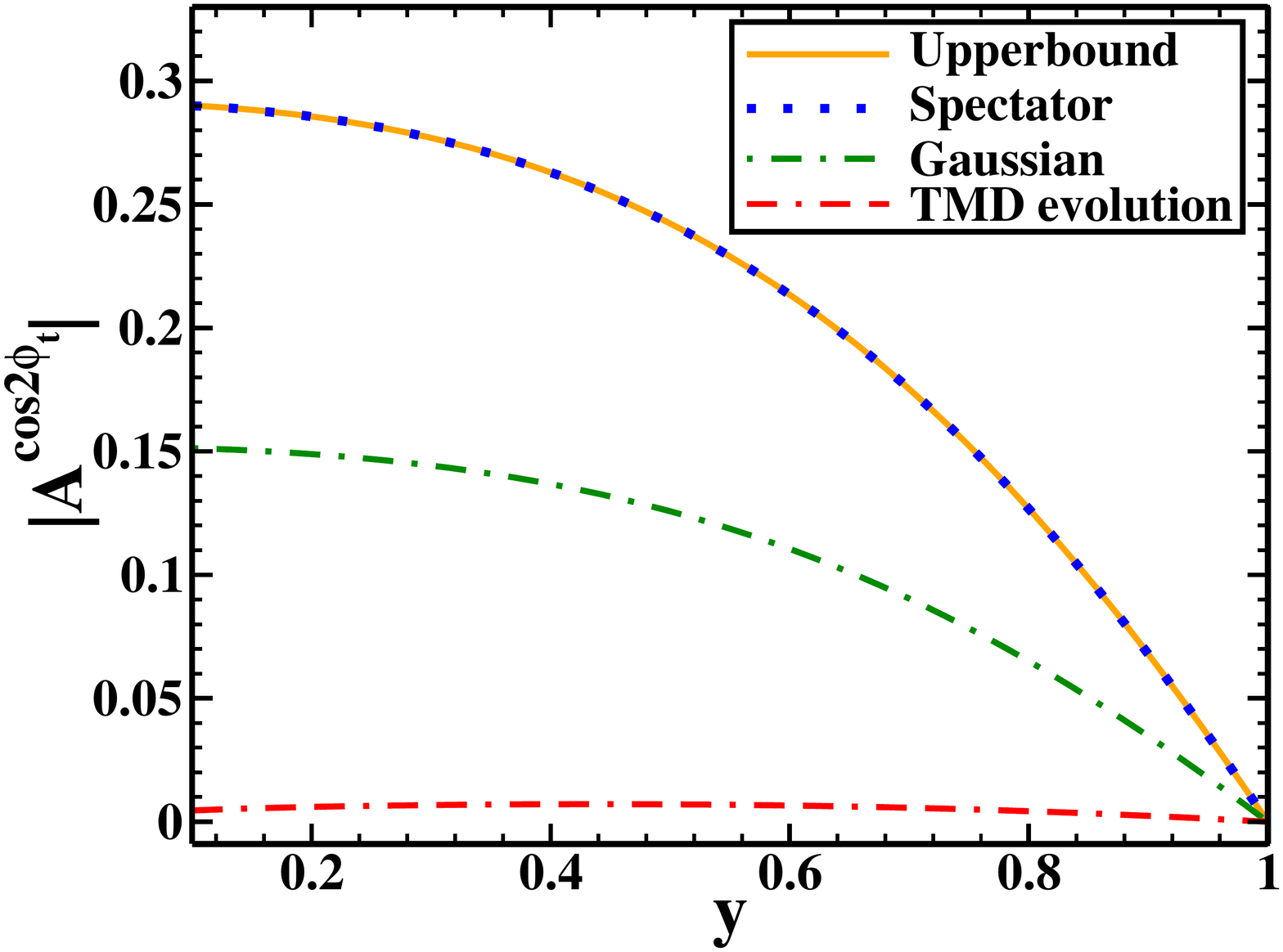}
				\hspace{0.1cm}
				\small{(d)}\includegraphics[width=8cm,height=6.5cm,clip]{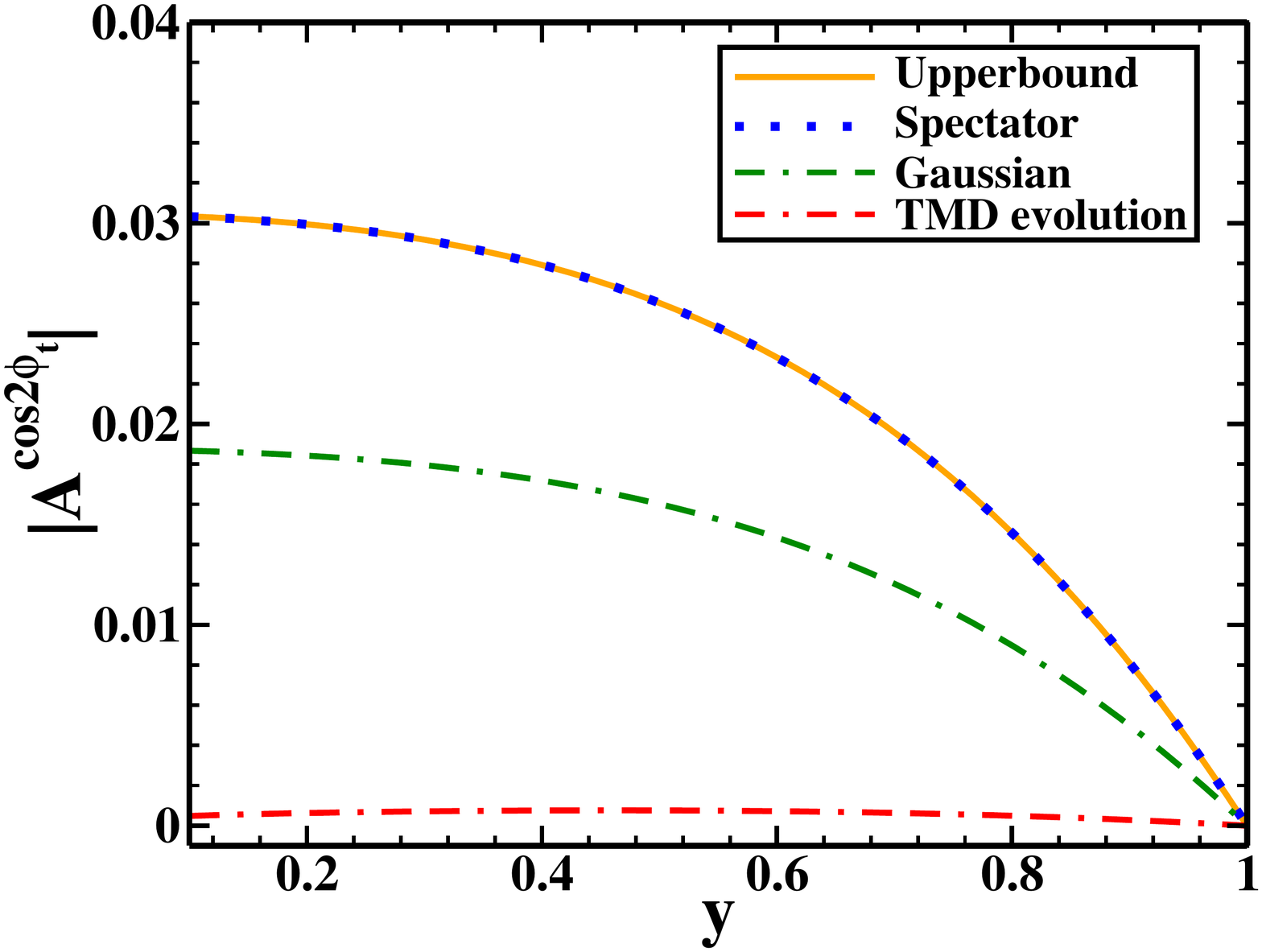}
			\end{minipage}
			\caption{ Upper bound of the asymmetry compared with the absolute values of $A^{\cos2\phi_t}$ calculated in spectator model, Gaussian model and TMD evolution, respectively,  for $e^- + P \to e^-  + J/\psi  + Jet + X$ at $\sqrt{s}=140$.
			Left panel shows the asymmetry in NRQCD and right panel in CS. We have taken $y=0.3$ in the upper panel ((a) and (b)) and  $\mathrm{K}_t=2$ GeV in the lower panel ((c) and (d)).}
			\label{comp}
		\end{figure}

In Figs.~\ref{evol}-\ref{gauss}, we show a comparison of the $\cos2\phi_t$ azimuthal asymmetry using three different models/parameterizations of the gluon TMDs. Later, in Fig.~\ref{comp}, we have compared them with the asymmetry calculated by satisfying the upper bound of the TMDs, Eq. (\ref{eq:ub}). In all plots, we have shown the results when only the CS contributions are included(CS), as well as when both CS and CO contributions are included in NRQCD (NRQCD).   

In Fig. \ref{evol}, we plot the $\cos2\phi_t$ azimuthal asymmetry in the TMD evolution approach as a function of $\mathrm{K}_t$, $y$ and $z$ at the center of mass energy $\sqrt{s}= 140~$GeV.  The integration ranges are $\mathrm{q}_t\in [ 0.0-1.0 ]$  and  $y\in [0.1-1.0]$. The range of $\mathrm{q}_t$ is considered to satisfy the condition $|\bm{\mathrm{q}}_t|\ll |\bm{\mathrm{K}}_t|$.  A similar set of plots is shown for the spectator model  and for the Gaussian parameterization of the gluon TMDs in Fig.~\ref{spec} and \ref{gauss} respectively.
In all these plots, we see that in contrast to the CS case, the NRQCD framework gives a significant contribution to $\cos2\phi_t$  azimuthal asymmetry at $\sqrt{s}=140~$GeV.  The magnitude of the asymmetry does not change that much if we take a somewhat lower value of $\sqrt{s}$ for example $\sqrt{s} = 65 $ GeV. 


In the upper left panel (a) of Figs.~\ref{evol}-\ref{gauss}, we have plotted the $\cos2\phi_t$ asymmetry as functions of $\mathrm{K}_{t}$ at $\sqrt{s}=140~$GeV. In these plots, we integrated $\mathrm{q}_t$ and $y$ in the range  $(0,1)$ and, $(0.1,1)$ respectively. We see that the $\cos2\phi_t$ asymmetry is  maximum (negative)  for lower $\mathrm{K}_t$ and monotonically decreases as we go in the higher $\mathrm{K}_t$ region. The maximum asymmetry we obtained is $\approx 29\%$ in the spectator model at $\mathrm{K}_t=1~$GeV followed by the Gaussian parameterization, $\approx 17\%$. Incorporation of the  TMD evolution results in  a smaller  asymmetry, $\approx 1\%$ at $\mathrm{K}_t=1$ GeV.

The $y$ dependence of $\cos2\phi_t$ azimuthal asymmetry is shown in the upper right panel (b) of Figs.~\ref{evol}-\ref{gauss} at $\sqrt{s}=140~$GeV; we have shown the results both in  NRQCD and in the CS. We plotted the asymmetry for two fixed values of $\mathrm{K}_t$, namely  $2~$GeV and $4~$GeV. We have plotted the asymmetry in the range of $y\in [0.1,1]$, however in the lower $y$ region, the magnitude of the asymmetry is similar in both the spectator and the Gaussian models, whereas in TMD evolution approach, the asymmetry is maximum around $y=0.44$ at $\mathrm{K}_t=2~$GeV in NRQCD. The asymmetry is small in the TMD evolution approach;  however, we obtain a significant asymmetry,  $\approx 29\%$, in the spectator model followed with $\approx 18\%$ in the Gaussian model at $\mathrm{K}_t=2~$GeV and $y=0.1$.

In all the above discussed plots of the asymmetry as functions of $\mathrm{K}_t$ and $y$, we have taken a fixed value of $z=0.7$. However, in the lower panel (c) of Figs.~\ref{evol}-\ref{gauss}, we have plotted the $\cos(2\phi_t)$ asymmetry as functions of $z$ at $\sqrt{s}=140~$GeV and $\mathrm{K}_t=3~$GeV for both NRQCD and the CS. We have taken fixed values of $y$, namely,  $0.1$ and $0.8$. The peak of the asymmetry is $\approx 12\%$  at $z\approx 0.7$ in NRQCD and $\approx 2\%$ at $z\approx 0.4$ in the CS at $y=0.1$. In all these plots, we have used LDMEs from \cite{ChaoLDME}.


\begin{figure}[H]
			\begin{minipage}[c]{0.91\textwidth}
				\small{(a)}\includegraphics[width=8cm,height=6.5cm,clip]{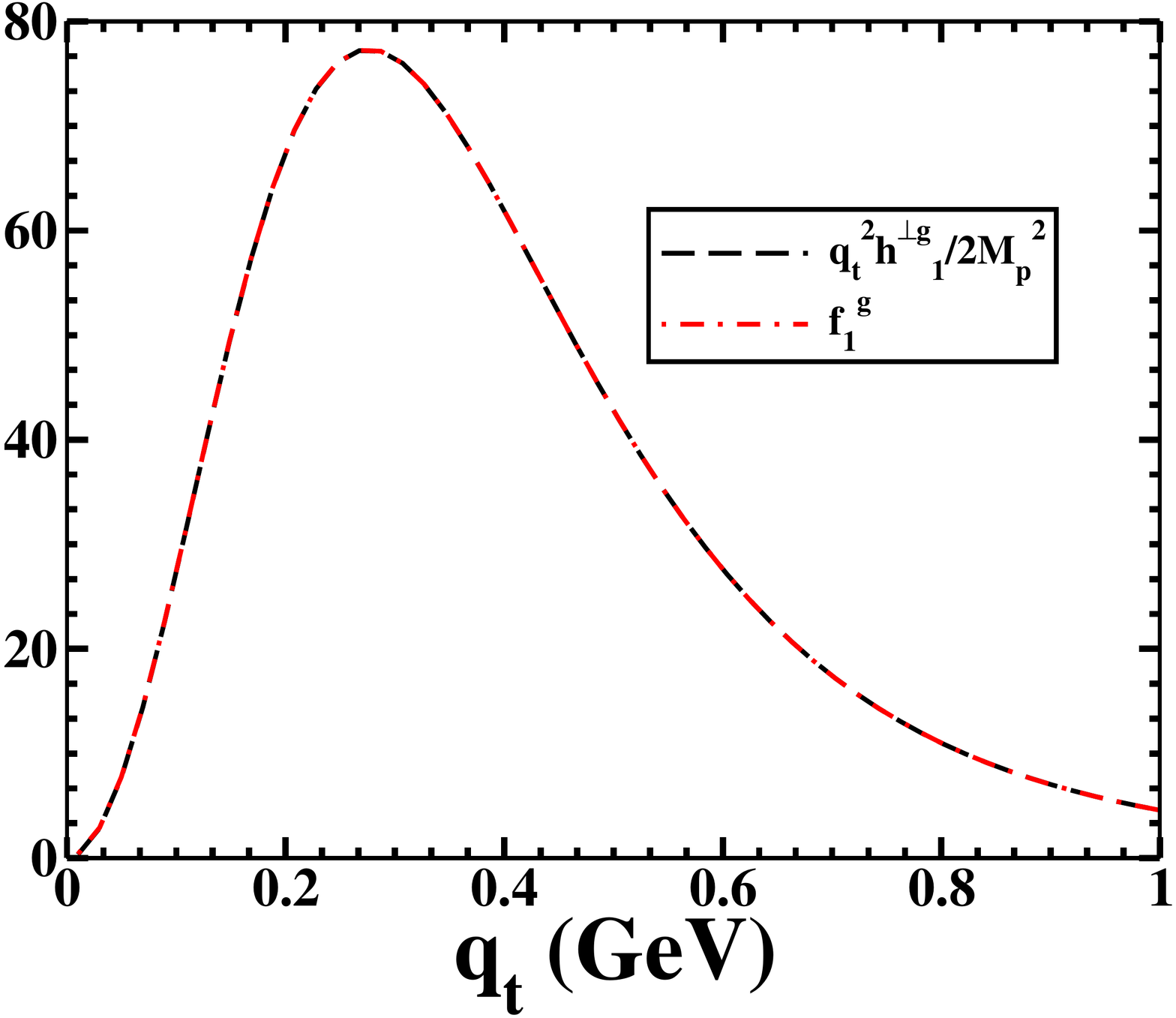}
				\hspace{0.1cm}
				\small{(b)}\includegraphics[width=8cm,height=6.5cm,clip]{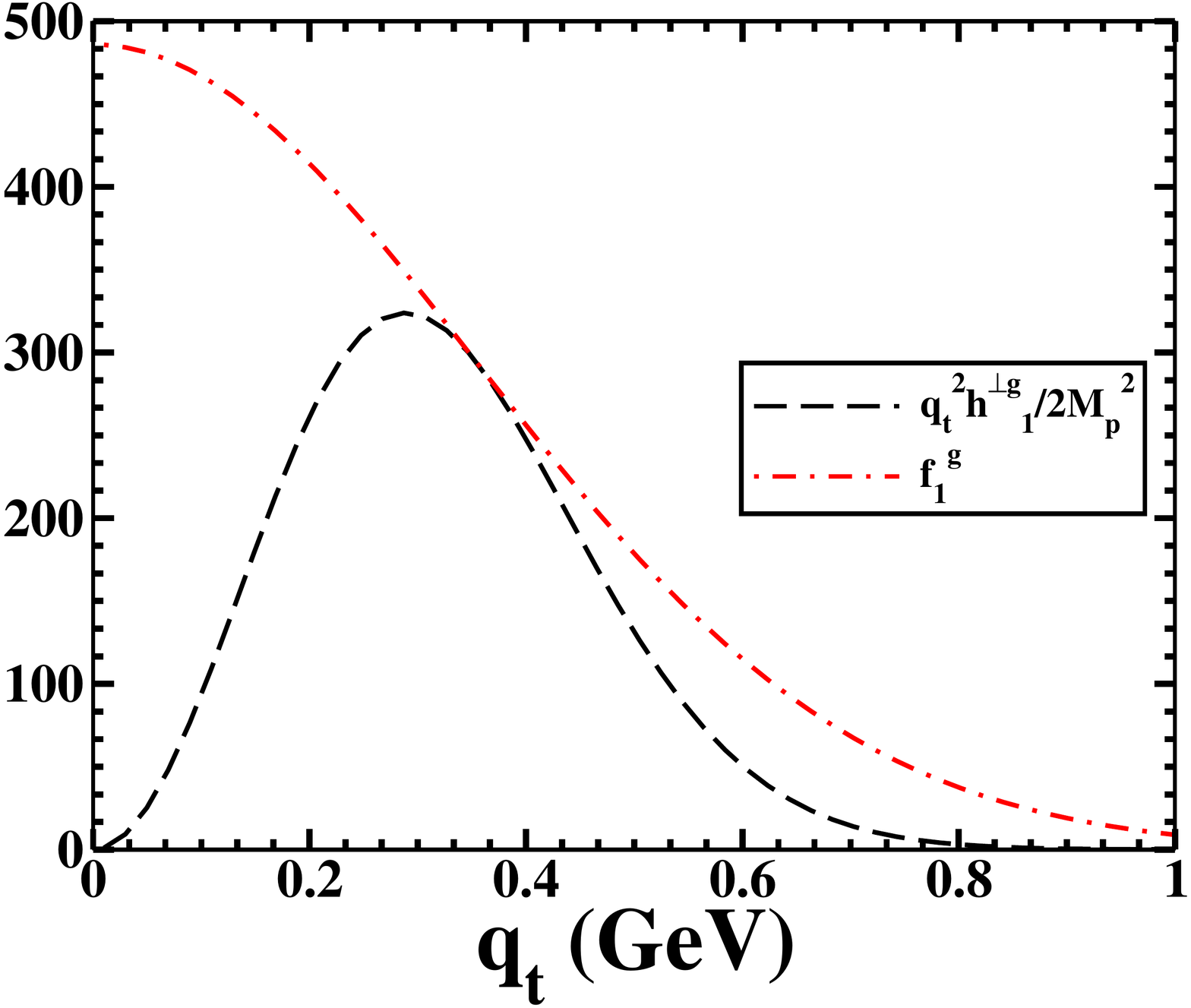}
			\end{minipage}
			\begin{minipage}[c]{0.91\textwidth}
				\small{(c)}\includegraphics[width=8cm,height=6.5cm,clip]{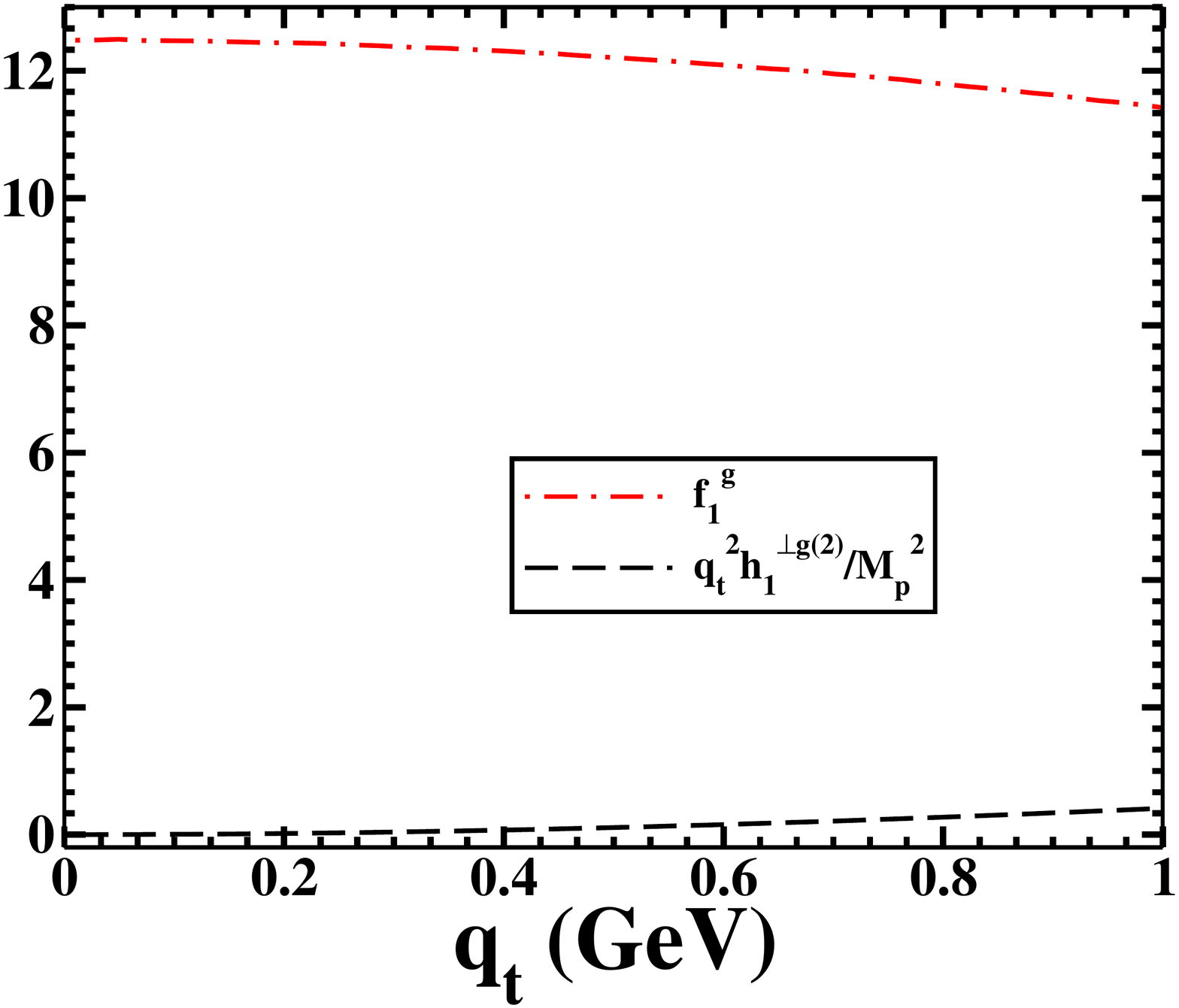}
				\hspace{0.1cm}
				\small{(d)}\includegraphics[width=8cm,height=6.5cm,clip]{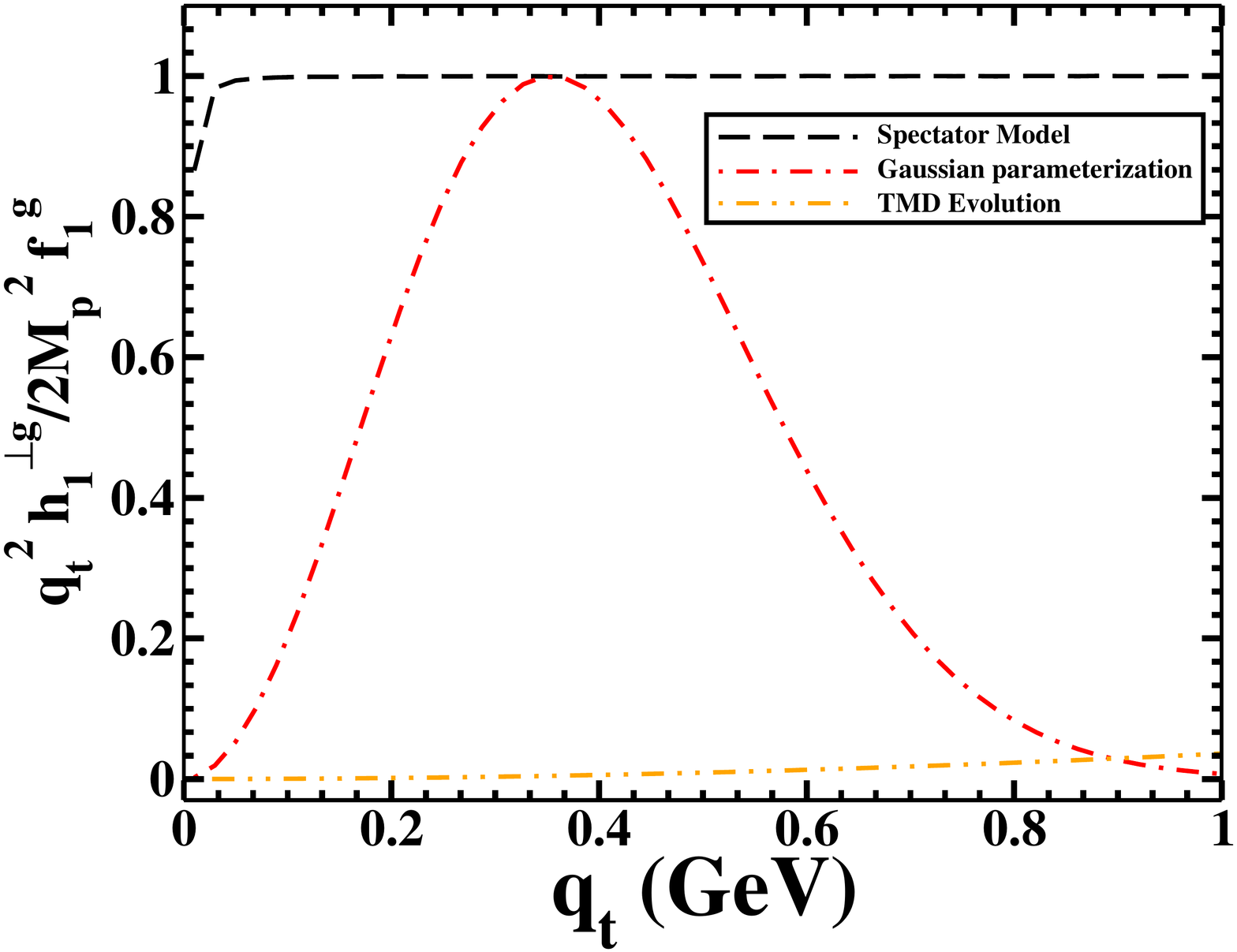}
			\end{minipage}
			\caption{Unpolarized and linearly polarized gluon TMDs as a function of $\mathrm{q}_t$ calculated in spectator model (a), Gaussian model (b) and TMD evolution (c), respectively at $\sqrt{s}=140$, $\mathrm{K}_t=3.0\mathrm{GeV}$, $y=0.3$ and $z=0.7$. (d) panel shows the comparison between the positivity bound for all the parameterizations.}
			\label{plot:TMDs}
\end{figure}

In Fig.~\ref{plot:TMDs}, we have plotted both the TMDs, $f_1^g$ and $h_1^{\perp g}$ and their ratio, $\frac{\mathrm{q}_t^2h_1^{\perp g}}{2M_P^2 f_1^g}$ as functions of $\mathrm{q}_t$ for all three parameterizations. In all these plots we have used similar kinematics as we considered for the plots in Fig.~\ref{evol}-\ref{comp}. In this kinematics, the $x$ value of the gluon TMDs are of the order of $10^{-3}-10^{-2}$.  We have plotted at the probing scale which is the virtuality of photon, $Q^2=M_\psi^2+\mathrm{K}_t^2$, where $\mathrm{K}_t=3$ GeV and at the fixed values of $y=0.3$ and $z=0.7$. This sets  $x\approx 0.012$. From the plots of the ratio, $\frac{\mathrm{q}_t^2h_1^{\perp g}}{2M_Pf_1^g}$ ((d) of Fig.~\ref{plot:TMDs}), we see  that the TMDs in the spectator model indeed saturates the positivity bound, whereas, the Gaussian parameterizations and TMD evolution approach satisfy the positivity bound but do not saturate it except for $\mathrm{q}_t \approx 0.36$ GeV, where Gaussian parameterization is saturating the positivity bound. Moreover, the ratio is larger in the case of Gaussian as compared with TMD evolution approach for almost whole range of $\mathrm{q}_t$ considered. In the Spectator model, tails of the TMDs in the small-$x$ domain, depend on the trend of spectral function at large $M_X$ \cite{Bacchetta:2020vty,Bacchetta:2020vty}. We have checked that the  spectator model results in Eqs. (12) and (15) of \cite{Bacchetta:2020vty} for the ratio  $\frac{\mathrm{q}_t^2h_1^{\perp g}}{2M_P^2 f_1^g}$ at $x=0.001$ without integrating over the spectral function, does not saturate the positivity bound when $M_X$ is large. However, if we multiply by the spectral function and integrate over $M_X$,  the TMDs saturate the positivity bound when $x$ is of the order of $10^{-3}-10^{-2}$; this could be  because the spectral function is zero for higher values of $M_X$ in replica 11. However, in the higher $x$ region, the TMDs  do not saturate the bound but satisfies it for the whole range of the transverse momentum, $\mathrm{q}_t$.  


In Fig. \ref{comp}, we show a comparison of the upper bound of the asymmetry with that calculated in Spectator model, Gaussian model and TMD evolution approach at $\sqrt{s}=140$ GeV. The upper bound of the asymmetry is  calculated by saturating the  positivity bound of TMDs in Eq. (\ref{eq:ub}) and fixing all the parameters mentioned above. We have shown the result both in  NRQCD and CS  as a function of $\mathrm{K}_t$  at $y=0.3$ (upper panel) and as a function of $y$ at $\mathrm{K}_t=2$ GeV in the lower panel. 
In case of TMD evolution, the non - perturbative Sudakov factor corresponding to $\mathrm{b}_{tlim}=2~\mathrm{GeV}^{-1}$. For all the plots the range of integration of $\mathrm{q}_t\in [0.0-1.0] \mathrm{GeV}$, $z=0.7$ and we have fixed the virtuality of photon $Q=\sqrt{M_\psi^2+\mathrm{K}_t^2}$.  

We can see that the asymmetry calculated with the spectator model is maximum and agrees with  the upper bound. As seen from Fig. \ref{comp}, the asymmetry incorporating TMD evolution is significantly smaller than that calculated using  Gaussian and spectator models for the gluon TMDs. This is because the denominator of the asymmetry receives contribution from the unpolarized gluon distribution which has a leading order (LO) term Eq. (\ref{f1g}) whereas the numerator contains the linearly polarized gluon distribution whose leading contribution comes at $O(\alpha_s)$. If we exclude the LO term in the unpolarized gluon TMD, we find the asymmetry increases approximately three  times. 
We also find that the magnitude of the asymmetry does not change much if $\sqrt{s}$ is lower.


\begin{figure}[H]
			\centering
			\subfigure[]{\includegraphics[width=8cm,height=6.5cm,clip]{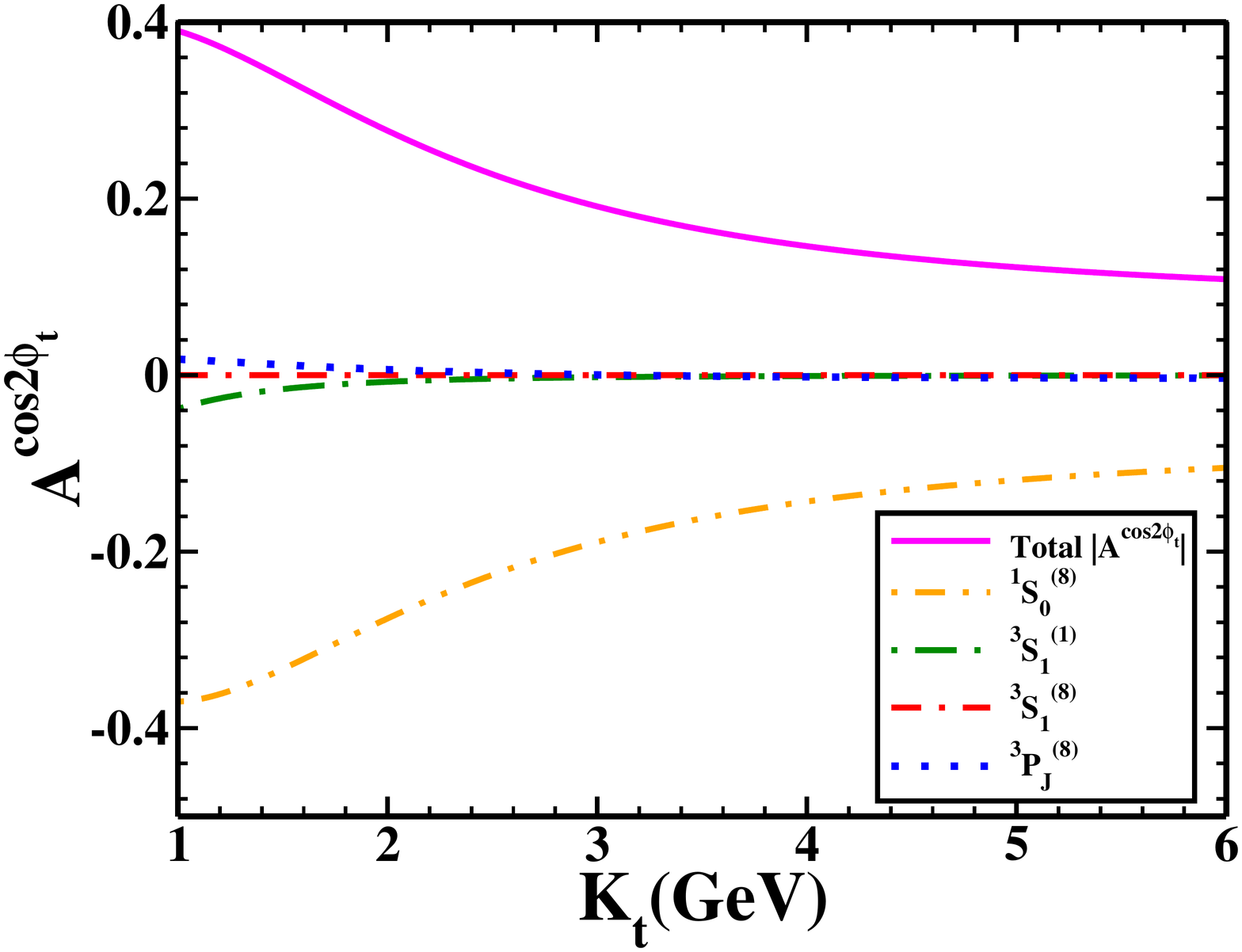}}
			\subfigure[]{\includegraphics[width=8cm,height=6.5cm,clip] 	{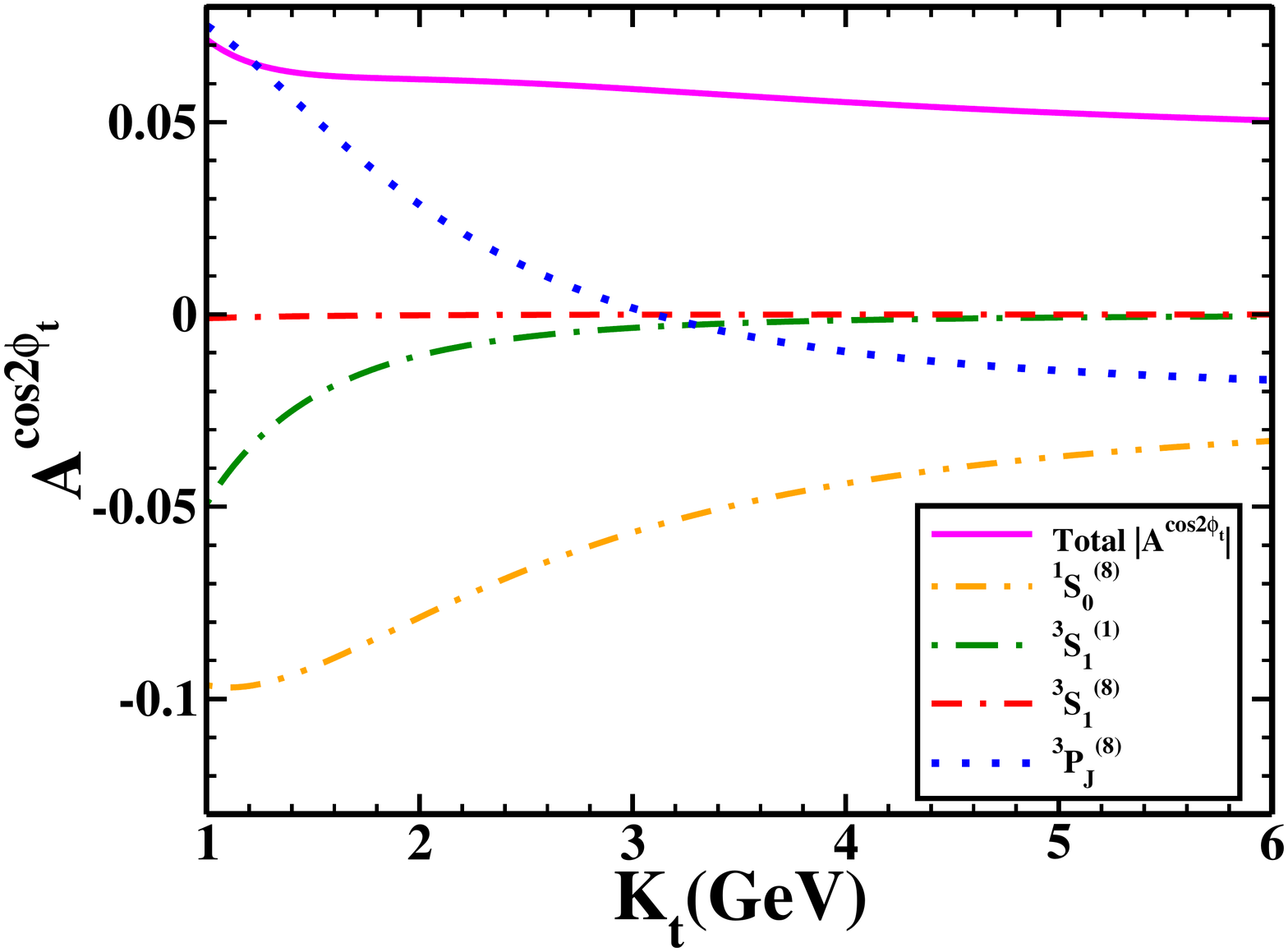}}
			\caption{Contribution to the upper bound of the asymmetry coming from individual states, as a function of $\mathrm{K}_t$ at $\sqrt{s}=140$ GeV, $Q=\sqrt{M_{\psi}^2+\mathrm{K}_t^2}$ and $y=0.3$; (a)  using the CMSWZ set of LDMEs \cite{ChaoLDME} and (b) using the SV set of LDMEs \cite{SharmaLDME}}
			\label{UPPER}
		\end{figure}

Lastly, in Fig. \ref{UPPER}, we show the upper bound for the absolute value of $|A^{\cos2\phi_t}|$ within the NRQCD using two sets of LDMEs,  as well as the contributions coming from individual states $i.e.$, $^1S_0^{(8)}, ^3S_1^{(1,8)}$ and $^3P_j^{(8)}$. One can see from (a) that for the LDME set CMSWZ \cite{ChaoLDME}, the dominating contribution comes from one single state, $^1S_0^{(8)}$; while from (b) one can see that for SV set of LDME \cite{SharmaLDME}, the dominating contribution  comes from two states, $^1S_0^{(8)}$ and $^3P_j^{(8)}$. So, we can conclude that the asymmetry depends on the LDME set chosen. It is worth mentioning here that our results for the upper bound of the asymmetry do not match with those presented in \cite{DAlesio:2019qpk} even if we plot it using the same scale as in this reference. We have traced this mismatch to a difference in the sign of the contribution coming from the $^3P_0^{(8)}$ state to the coefficient $\mathbb{B}_0$.

\section{Conclusion}
We have presented a calculation of the $\cos2\phi_t$ asymmetry in almost back-to-back production of a $J/\psi$ and a jet in $ep$ collision, using TMD factorization and generalized parton model. This asymmetry is sensitive to the still unknown linearly polarized gluon distribution. We present a numerical estimate of the asymmetry in the kinematical region that will be accessible at the future EIC. We have used NRQCD to calculate the $J/\psi$ production rate and two recent parameterization for the gluon TMDs, one based on a Gaussian type distribution and another based on spectator model. The asymmetry is quite sizable; in fact in spectator model the asymmetry agrees with the upper bound that is obtained by saturating the positivity condition of the gluon TMDs. TMD evolution affects the asymmetry at the energy of the EIC, making it smaller. The asymmetry also depends on the LDMEs used, and dominating contribution comes from different states. We conclude that the back-to-back production of  $J/\psi$ and a jet at the future EIC will be a very useful channel to probe the linearly polarized gluon TMDs.   

\section{Acknowledgement}
We acknowledge the funding from Board of Research in Nuclear Sciences (BRNS), Govt. of India,  under sanction No. 57/14/04/2021-BRNS/57082. We thank A. Bacchetta, M. Radici and F. Celiberto for useful discussion.

		\bibliographystyle{apsrev}
		\bibliography{biblography_as.bib}
		
	\end{document}